%% file: main.tex
\documentclass[a4paper,11pt]{article}
\usepackage{jcappub} 
\usepackage{lineno}
\usepackage{xcolor}
\usepackage{multirow}
\usepackage{array}
\usepackage{graphics}
\usepackage{hhline}
\usepackage{makecell}
\usepackage{arydshln}
\usepackage{orcidlink}
\usepackage{caption}
\usepackage{subcaption}
\usepackage{graphicx}
\usepackage{soul}
\usepackage[capitalise,nameinlink,noabbrev]{cleveref}

\newcommand{\hmpcinv}{\,h\,{\rm Mpc^{-1}}}
\newcommand{\hinvmpc}{\,h^{-1}{\rm Mpc}}
\newcommand{\hinvMpccubed}{\, h^{-3} \, \text{Mpc}^{3}}
\newcommand{\Hquad}{\hspace{0.5em}} 
\newcommand{\truncnorm}[4]{$\mathcal{N}_{[#1,#2]}(#3, #4^2)$}
\newcommand{\leftparbox}[2]{\parbox{#1}{\begin{flushleft} #2 \end{flushleft}}}

\makeatletter
\input{aas_macros.sty}
\let\jnl@style=\relax
\makeatother

\title{Enhancing DESI DR1 Full-Shape analyses using HOD-informed priors}

\affiliation{Affiliations are in \cref{sec:affiliations}}
\input{author}

\emailAdd{hanyu.zhang@uwaterloo.ca}

\abstract{We present an analysis of DESI Data Release 1 (DR1) that incorporates Halo Occupation Distribution (HOD)-informed priors into Full-Shape (FS) modeling of the power spectrum based on cosmological perturbation theory (PT). By leveraging physical insights from the galaxy–halo connection, these HOD-informed priors on nuisance parameters substantially mitigate projection effects in extended cosmological models that allow for dynamical dark energy. The resulting credible intervals now encompass the posterior maximum from the baseline analysis using gaussian priors, eliminating a significant posterior shift observed in baseline studies. In the $\Lambda$CDM framework, a combined DESI DR1 FS information and constraints from the DESI DR1 baryon acoustic oscillations (BAO)—including Big Bang Nucleosynthesis (BBN) constraints and a weak prior on the scalar spectral index—yields $\Omega_{\rm m} = 0.2994\pm 0.0090$ and $\sigma_8 = 0.836^{+0.024}_{-0.027}$, representing improvements of approximately 4\% and 23\% over the baseline analysis, respectively. For the $w_0w_a$CDM model, our results from various data combinations are highly consistent, with all configurations converging to a region with $w_0 > -1$ and $w_a < 0$. This convergence not only suggests intriguing hints of dynamical dark energy but also underscores the robustness of our HOD-informed prior approach in delivering reliable cosmological constraints.}

\begin{document}
\maketitle
\flushbottom

\section{Introduction}
\label{sec:intro}

Observations of large-scale structure (LSS) are a cornerstone of modern cosmology, offering critical insights into the growth of cosmic structures, the nature of dark matter and dark energy and the initial conditions of the Universe. These measurements can be realized through the spatial distribution of galaxies, encoded by their clustering statistics in comoving space and their projection into observed galaxy angles and redshifts. Previous-generation surveys, Baryon Oscillation Spectroscopic Survey (BOSS) \cite{Dawson2013TheSDSS-III} and its extension eBOSS \cite{Dawson2016TheData}
have successfully demonstrated the power of LSS measurements in revealing key cosmological information, yielding constraints that support the $\Lambda$CDM concordance cosmological model \cite{Alam2021CompletedObservatory}.

The Dark Energy Spectroscopic Instrument (DESI), the first Stage-IV galaxy survey in operation \cite{Levi2013The2013,DESICollaboration2016TheDesign,Collaboration2024ValidationInstrument,Collaboration2024TheInstrument}, is undertaking a five-year spectroscopic program covering 14,200 square degrees of the sky \cite{DESICollaboration2016TheDesignb, Collaboration2022OverviewInstrument,Silber2023TheDESI,Miller2024TheInstrument,Schlafly2023SurveyInstrument,Guy2023TheInstrument,FiberSystem.Poppett.2024}. The survey targets five specific tracers—Bright Galaxy Survey (BGS) \cite{Hahn2023TheValidation}, luminous red galaxies (LRG) \cite{Zhou2023TargetGalaxies}, emission line galaxies (ELG) \cite{Raichoor2023TargetGalaxies}, quasars (QSO) \cite{Chaussidon2023TargetQuasars}, and the Ly$\alpha$ forest \cite{Myers2023TheInstrument}—spanning a redshift range of $0 < z < 4$. In total, DESI will collect precise redshifts for approximately 50 million galaxies and quasars, providing an unprecedented view of the LSS of the Universe. The DESI Data Release 1 (DR1) \cite{2025arXiv250314745D} contains spectra from 1 year of regular observations, and recent science results using the DESI DR1 data have further advanced our understanding of cosmology through the measured galaxy clustering signal \cite{DESICollaboration2024DESIQuasars,Adame2025DESIOscillations,Adame2025DESIForest,DESICollaboration2024DESIQuasarsb,DESICollaboration2024DESIMeasurements, Ishak2024Modified2024}. Part of this is through the measurement of the baryon acoustic oscillations (BAO). This series of peaks and troughs in the clustering power spectrum is used as a robust standard ruler to measure cosmological expansion \cite{Eisenstein1998CosmicSurveys,Blake2003ProbingRuler,Seo2003ProbingSurveys}. Going beyond the BAO signal, we need to measure and model the Full-Shape clustering of DESI tracers. This makes more demands on the models, with the potential rewards that Full-Shape clustering not only captures the signal of cosmic structure growth but also encodes crucial information about the amplitude and shape of the primordial power spectrum \cite{Peebles1980TheUniverse, Liddle2000CosmologicalStructure, Koyama2016CosmologicalGravity, Joyce2016DarkGravity, Ishak2019TestingCosmology, Alam2021TowardsRequirements, Huterer2023GrowthStructure}. 

The standard theoretical framework commonly adopted for interpreting Full-Shape measurements is based on cosmological perturbation theory (PT) \cite{Bernardeau2002Large-scaleTheory}, which decomposes models of the non-linear evolution of the matter power spectrum into a series of terms dependent on increasing powers of the overdensity. Recently, this framework was augmented with effective field theory (EFT) techniques that incorporate a series of counterterms to capture various small-scale effects \cite{Baumann2012CosmologicalFluid,Carrasco2012TheStructures,Porto2014TheStructures,Perko2016BiasedStructure, Lewandowski2017AnRegime,Ivanov2022EffectiveStructure}. The EFT approach systematically accounts for non-linear gravitational dynamics and other complex processes influencing galaxy formation and distribution, proving instrumental in translating recent observations of galaxy clustering into robust cosmological constraints \cite{Colas2020EfficientStructure,dAmico2020TheStructure,Ivanov2020CosmologicalSpectrum,DAmico2021LimitsCode,Niedermann2021NewData,Kumar2022UpdatingGalaxies, Simon2022ConstrainingStructures,Nunes2022NewSpectrum,Philcox2022BOSSMonopole,Zhang2022BOSSStructure,Chen2022ABAO,Lague2022ConstrainingSurveys,Simon2023ConsistencySpectrum,Carrilho2023CosmologyPriors,Schoneberg2023ComparativeRadiation,Smith2023AssessingConstant,Allali2023DarkDatasets,Simon2023CosmologicalAnalysis,Simon2023UpdatedEnergy,DAmico2024LimitsEFTofLSS}.
However, this approach introduces a number of nuisance parameters—controlling the galaxy bias, counter terms, and shot noise contributions—that must be marginalized over in a Bayesian analysis.
These nuisance parameters are often degenerate with the cosmological parameters of interest; when coupled with weakly constrained or poorly defined priors on them, they can lead to projection effects \cite{Simon2023ConsistencySpectrum}. Such effects arise when unconstrained regions of the nuisance parameter space contribute disproportionately to the marginalized posterior of the cosmological parameters, resulting in a marginalized posterior that deviates from the Maximum a Posterior (MAP) value—especially in extended cosmological models \cite{Simon2023ConsistencySpectrum,Gomez-Valent2022FastCosmology,Carrilho2023CosmologyPriors, Hadzhiyska2023CosmologyParameters, DAmico2024TheStructure, Maus2025AnBeyond}.
In baseline analyses \cite{DESICollaboration2024DESIMeasurements}, Gaussian priors are employed for most nuisance parameters, with their widths determined from extensive validation tests against simulations \cite{DESICollaboration2024DESIQuasarsb,Maus2025AnBeyond}. Nevertheless, due to the inherently weak constraints on these parameters, the baseline approach still suffers from projection effects.
Several strategies have been explored to mitigate this issue, including the use of simulation-based priors for nuisance parameters \cite{Zhang2024HOD-informedLSS,Ivanov2024Full-ShapeBOSS,Ivanov2024Full-ShapeAnomaly} (with \cite{ 2025arXiv250307270I} proposing an analytic approach to generate approximate simulation‐based priors), non-linear re-parameterization to de-correlate nuisance parameters from the cosmological parameters \cite{Paradiso2024ReducingStructure,desi-gam}, and frequentist inferences that do not explicitly rely on prior assumptions \cite{Holm2023DecayingLikelihoods,Holm2023BayesianData,Herold2024ProfileEnergy,desi-frequentist}. 

In this work, we address these challenges by employing physically motivated priors for DESI tracers that restrict the nuisance parameter space, using the method outlined in \cite{Zhang2024HOD-informedLSS}. The priors we adopt are grounded in the Halo Occupation Distribution (HOD) framework \cite{Jing1998SpatialSurvey,Peacock2000HaloBias,Seljak2000AnalyticClustering,Scoccimarro2001HowClustering,Cooray2002HaloStructure,Berlind2002TheMass,Zheng2005TheoreticalGalaxies,Zheng2007GalaxyClustering,Zheng2009HaloGalaxies}, ensuring that nuisance parameters are restricted to models consistent with a physical galaxy–halo connection.
This enhances the robustness and precision of our cosmological inference and mitigates the severe projection effects observed in certain model–dataset combinations seen in the baseline DESI DR1 Full-Shape analysis \cite{DESICollaboration2024DESIQuasarsb, DESICollaboration2024DESIMeasurements}. 

The remainder of the paper is organized as follows: in \cref{sec:fs_hip}, we describe the data and modeling framework for DESI Full-Shape measurements and the construction of HOD-informed priors. \cref{sec:likelihoods_inference} outlines our likelihood and cosmological inference methodology. In \cref{sec:results}, we present our results for both the $\Lambda$CDM and $w_0w_a$CDM models. Finally, we summarize our findings and discuss their implications for future research in \cref{sec:conclusions}.

\section{DESI Full-Shape and HOD-informed priors}
\label{sec:fs_hip}
We adopt a perturbation theory–based method that directly fits a model to the measured Full-Shape power spectrum multipoles, we briefly review the Full-Shape measurements and modeling before describing our construction of HOD-informed priors for the nuisance parameters—an approach designed to reconcile projection effects and guide the parameter space in a physically motivated manner \cite{Zhang2024HOD-informedLSS}.

\subsection{DESI Full-Shape measurements and modeling}
\label{subsec:fs}
\subsubsection{Full-Shape measurements}
\label{subsubsec:fs_meassurements}
DESI DR1 encompasses over 4.7 million redshift measurements spanning the redshift range $0.1<z<2.1$, split into separate target classes. Bright Galaxy Survey (BGS) galaxies cover $0.1<z<0.4$, Luminous Red Galaxies (LRGs) are split into three redshift bins ($0.4<z<0.6$, $0.6<z<0.8$, and $0.8<z<1.1$), Emission Line Galaxies (ELGs) span $1.1<z<1.6$, and Quasars (QSOs) cover $0.8<z<2.1$ (see \cite{DESICollaboration2024DESIStatistics} for further details on the sample selection and characteristics).

In order to compare our results with those in the baseline analysis \cite{DESICollaboration2024DESIQuasarsb, DESICollaboration2024DESIMeasurements}, we match as closely as possible the analysis and model fitting method of that work. We extract power spectrum multipole measurements from the DESI tracers using the Feldman-Kaiser-Peacock (FKP) \cite{Feldman1994Power-SpectrumSurveys} estimator \cite{Yamamoto2006ASurvey, Bianchi2015MeasuringFFTs, Hand2018Nbodykit:Structure} implemented in the \texttt{pypower} code \footnote{\url{https://github.com/cosmodesi/pypower}}. Galaxies are weighted to account for the selection function and to optimally measure two-point statistics (see \cite{DESICollaboration2024DESIStatistics} for a detailed description of the codes and weighting scheme). The small-scale signal imprinted by DESI fiber assignment is mitigated through a combination of the $\theta$-cut method \cite{Pinon2025MitigationEstimators} and a rotation of the data vector, window matrix, and covariance, which helps achieve a more diagonal window function \cite{DESICollaboration2024DESIStatistics}. The covariance matrix is estimated from 1000 EZmocks and then rescaled to match the semi-empirical covariance predicted from the observed data \cite{Forero-Sanchez2024AnalyticalResults, Rashkovetskyi2025Semi-analyticalData,KP4s8-Alves,KP3s8-Zhao}. Contributions from various systematic effects are directly incorporated into the covariance. One of the components of these systematic effects is the prior weight effect. In \cite{Findlay2024ExploringAnalysis}, the prior weight effect was quantified for a variety of HOD models using the baseline prior. However, when using the HOD-informed prior, the contribution from the prior weight effect is expected to be reduced. Since this reduction is not precisely quantified, we choose to adopt a conservative approach by continuing to include the additional contribution to the covariance suggested by \cite{Findlay2024ExploringAnalysis}. In our analysis, we focus on the monopole and quadrupole measured over the wavenumber range $0.02<k<0.2 \hmpcinv$ with a binning width of $\Delta k=0.005 \hmpcinv$, in line with the baseline choices.

\subsubsection{PT-based modeling}
\label{subsubsec:fs_model}
\input{Tables/configs}
To model the measurements, we adopt a perturbation theory–based approach that directly fits the Full-Shape power spectrum multipoles. 
Specifically, we use the \texttt{desilike} \footnote{\url{https://github.com/cosmodesi/desilike}} framework, which employs the \texttt{velocileptors} \footnote{\url{https://github.com/sfschen/velocileptors}} \cite{Chen2020ConsistentTheory, Chen2021Redshift-spaceTheory} library for theoretical modeling. In this analysis, we select the one-loop Eulerian Perturbation Theory (EPT) option from \texttt{velocileptors} to model the redshift-space galaxy power spectrum.
This framework incorporates additional counterterms that capture the impact of small-scale physics—such as galaxy formation processes—and includes stochastic contributions to account for shot noise and Fingers-of-God effects. The theoretical power spectrum is properly infrared resummed to handle long-wavelength modes that could otherwise lead to divergences in perturbative calculations \cite{Senatore2014RedshiftStructures, Senatore2015TheStructures, Lewandowski2020AnPeak}. As shown in \cite{Maus2025ASurveys}, the choice of PT code is not expected to impact the cosmological results. For further details on the PT model, we refer readers to \cite{DESICollaboration2024DESIQuasarsb}.

We adopt a set of variable parameters that follows the baseline analysis and is grounded in physical considerations, which helps align the model parameters more closely with the observed power spectrum multipoles and reduce biases from prior volume effects (see Appendix~B.2 of \cite{Maus2025AnBeyond} for details). Our variable parameter set is given by
\begin{align}
    \{b_{\rm 1p}, b_{\rm 2p}, b_{\rm sp}, b_{\rm 3p},\alpha_{\rm 0p},\alpha_{\rm 2p},\alpha_{\rm 4p},{\rm SN}_{\rm 0p},{\rm SN}_{\rm 2p},{\rm SN}_{\rm 4p}\}\,.
\end{align}
where $b_{\rm 1p}$, $b_{\rm 2p}$, $b_{\rm sp}$, and $b_{\rm 3p}$ are the galaxy bias parameters; $\alpha_{\rm 0p}$, $\alpha_{\rm 2p}$, and $\alpha_{\rm 4p}$ are the counterterm parameters; and ${\rm SN}_{\rm 0p}$, ${\rm SN}_{\rm 2p}$, ${\rm SN}_{\rm 4p}$ denote the stochastic parameters, with subscript ``p'' indicating that these quantities are defined in the physical basis. In practice, We set the third-order bias parameter $b_{\rm 3p}$ to zero, as it is expected to be small and degenerate with other nuisance parameters \cite{2014PhRvD..90l3522S,2018JCAP...09..008L}. Also since our analysis employs only the monopole and quadrupole, we fix $\alpha_{\rm 4p}$, and ${\rm SN}_{\rm 4p}$ to zero, as these parameters correspond to the hexadecapole contributions. These choices are consistent with the baseline DESI analysis. 

This physical basis can be converted into the Eulerian basis via
\begin{equation}
\begin{aligned}
&b_{\rm 1E} = \frac{b_{\rm 1p}}{\sigma_8}, \Hquad
b_{\rm 2E} = \frac{b_{\rm 2p}}{\sigma_8^2} + \frac{8}{21}\left(\frac{b_{\rm 1p}}{\sigma_8} - 1\right), \Hquad
b_{\rm sE} = \frac{b_{\rm sp}}{\sigma_8^2} - \frac{2}{7}\left(\frac{b_{\rm 1p}}{\sigma_8} - 1\right), \Hquad
b_{\rm 3E} = \frac{3b_{\rm 3p}}{\sigma_8^3} + \frac{b_{\rm 1p}}{\sigma_8} - 1, \\
&\alpha_{\rm 0E} = \left(\frac{b_{\rm 1p}}{\sigma_8}\right)^2 \alpha_{\rm 0p}, \Hquad
\alpha_{\rm 2E} = f\,\frac{b_{\rm 1p}}{\sigma_8}(\alpha_{\rm 0p}+\alpha_{\rm 2p}), \Hquad
\alpha_{\rm 4E} = f\left(f\,\alpha_{\rm 2p} + \frac{b_{\rm 1p}}{\sigma_8}\alpha_{\rm 4p}\right), \Hquad
\alpha_{\rm 6E} = f^2\,\alpha_{\rm 4p}, \\
&{\rm SN}_{\rm 0E} = {\rm SN}_{\rm 0p}\, \sigma^2_n,\quad
{\rm SN}_{\rm 2E} = {\rm SN}_{\rm 2p}\, \sigma^2_n\, f_{\rm sat}\, \sigma_v^2, \quad
{\rm SN}_{\rm 4E} = {\rm SN}_{\rm 4p}\, \sigma^2_n\, f_{\rm sat}\, \sigma_v^4.
\end{aligned}
\label{eqn:EPT_conversion}
\end{equation}
Here, the subscript ``E'' indicates parameters in the Eulerian basis. In these equations, $\sigma_8$ and $f$ represent the amplitude of mass fluctuations and the growth factor at the effective redshift of the tracer, respectively;  $\sigma_n^2$ denotes the tracer’s Poissonian shot noise; $f_{\rm sat}$ and $\sigma_v$ are the expected satellite fraction and the characteristic velocity for the tracer. The specific values are calibrated follow for each tracer are summarized in \cref{tab:config}. In particular, measurements of $z_{\rm eff}$ and $\sigma_n^2$ are detailed in \cite{DESICollaboration2024DESIStatistics}, while $f_{\rm sat}$ and $\sigma_v$ are calibrated following the methodology outlined in \cite{Maus2025AnBeyond}.
The resulting set of parameters, 
\begin{align}
\{b_{\rm 1E}, b_{\rm 2E}, b_{\rm sE}, b_{\rm 3E},\alpha_{\rm 0E},\alpha_{\rm 2E},\alpha_{\rm 4E},\alpha_{\rm 6E},{\rm SN}_{\rm 0E},{\rm SN}_{\rm 2E},{\rm SN}_{\rm 4E}\}
\end{align}
is then passed to \texttt{velocileptors} for theoretical calculations. 

\subsection{HOD-informed priors for DESI tracers}
\label{subsec:hip}
We now present our approach to construct HOD-informed priors (HIP) for DESI tracers, following the method outlined in \cite{Zhang2024HOD-informedLSS}. We begin by introducing the HOD model adopted for each DESI tracer, then describe the procedure used to derive the HIP, and finally present the resulting HIP for each tracer.

\subsubsection{HOD modeling of DESI tracers}
\label{subsubsec:hod_model}
The HOD model is a statistical framework that describes how galaxies occupy dark matter halos by specifying the probability distribution for the number of galaxies in a halo as a function of its properties. This approach links the dark matter distribution to the observed clustering of galaxies.  In this work, we adopt the form of the HOD models suggested by the DESI early data release \cite{Collaboration2024TheInstrument} for each tracer \cite{Yuan2024TheABACUSSUMMIT, Rocher2023TheSimulations, Smith2024GeneratingSurvey,Yuan2023UnravelingData}.

For LRGs, we adopt the HOD prescription described in \cite{Yuan2024TheABACUSSUMMIT}, originally proposed by \cite{Zheng2007GalaxyClustering}:
\begin{equation}
\begin{aligned}
    \langle N^{\rm LRG}_{\mathrm{cen}}\rangle(M) &= \frac{1}{2} \mathrm{erfc} \left(\frac{\log_\mathrm{10}(M_{\rm cut}/M)}{\sqrt{2}\sigma}\right)\,, \\
    \langle N^{\rm LRG}_{\mathrm{sat}}\rangle(M) &= \langle N_{\mathrm{cen}}\rangle(M)\left(\frac{M - \kappa M_{\rm cut}}{M_1}\right)^\alpha\,. \label{eqn:hod_lrg}
\end{aligned}
\end{equation}
In this model, the mean occupation of central and satellite galaxies differs: $M$ denotes the host halo mass, and $M_\mathrm{cut}$ is the minimum mass needed to host a galaxy. The parameter $\sigma$ controls the transition width for central galaxies probability with halo mass around $M_\mathrm{cut}$, $\kappa M_{\rm cut}$ sets the mass threshold for satellites, and the combination of $M_1$ and $\alpha$ determines how rapidly the number of satellites increases with halo mass.

For the BGS sample, we adopt the same HOD model as for LRGs. Since the BGS sample used in our Full-Shape analysis is defined by a fixed luminosity threshold, there is no need to incorporate luminosity evolution into the model, as described in \cite{Smith2024GeneratingSurvey}. 

For QSOs, we adopt an HOD model similar to that used for LRGs but remove the dependency on central modulation in the satellite occupation function \cite{Yuan2024TheABACUSSUMMIT}. This modification is motivated by the lack of evidence for a strong correlation between the presence of satellite QSOs and central QSOs.
\begin{equation}
\begin{aligned}
    \langle N^{\rm QSO}_{\mathrm{cen}}\rangle(M) &= \langle N^{\rm LRG}_{\mathrm{cen}}\rangle(M)\, ,\\
    \langle N^{\rm QSO}_{\mathrm{sat}}\rangle(M) &= \left(\frac{M - \kappa M_{\rm cut}}{M_1}\right)^\alpha\,. \label{eqn:hod_qso}
\end{aligned}
\end{equation}

Finally, for ELGs, we adopt the modified High Mass Quenched (mHMQ) model from \cite{Alam2020MultitracerEBOSS, Rocher2023TheSimulations}, using notation consistent with \cite{Yuan2022IllustratingILLUSTRISTNG,Yuan2023UnravelingData}. The model defines the expected number density of central galaxies as
\begin{equation}
\begin{aligned}
    \langle N^{\mathrm{ELG}}_{\mathrm{cen}}\rangle(M) &= p_{\max}\,\phi(M)\,\Phi(\gamma M),\\
    \phi(x) &= \frac{1}{\sqrt{2\pi}\,\sigma}
\,\exp\!\Biggl[-\,\frac{\bigl(x - \log_{10}(M_{\mathrm{cut}})\bigr)^2}{2\,\sigma^2}\Biggr],\\
    \Phi(x) &= \int_{-\infty}^{\,x}\phi(t)\,\mathrm{d}t \;=\; 1 + \mathrm{erf}\!\Biggl(\frac{x}{\sqrt{2}}\Biggr),
\end{aligned}
\label{eq:ELG_mHMQ}
\end{equation}
Here, $p_{\max}$ sets the maximum completeness level for central occupation. The function $\phi(x)$ is a Gaussian distribution with its peak defined by $\log_{10}(M_{\mathrm{cut}})$ and its width controlled by $\sigma$. The cumulative function $\Phi(\gamma M)$ introduces an asymmetry to the Gaussian via the tilt parameter $\gamma$. It is worth noting that while the same parameter name may have different physical interpretations in different HOD models, we retain a consistent nomenclature here for brevity. 

In the model, the satellite occupation function for ELGs largely follows a similar form to that used for QSO satellites, without any central modulation. This choice reflects the bursty nature of star formation that drives ELG selection, allowing for the possibility that ELG satellites reside in halos of lower mass than those hosting ELG centrals. In addition, several studies have found evidence that the ELG satellite occupation depends on the properties of the central galaxy \cite{Rocher2023TheSimulations, Yuan2023UnravelingData}. To incorporate this effect, we implement a central–satellite conformity prescription for ELG satellite \cite{Yuan2023UnravelingData}. Specifically, we define the satellite occupation function as
\begin{align}
\langle N^{\rm ELG}_{\mathrm{sat}}\rangle(M) &=
\begin{cases}
\left(\dfrac{M - \kappa\, M_{\rm cut}}{M_1}\right)^\alpha, & \text{if no ELG central}, \\[1mm]
\left(\dfrac{M - \kappa\, M_{\rm cut}}{M_{\rm 1EE}}\right)^\alpha, & \text{if ELG central}.
\end{cases}
\end{align}
This formulation allows the satellite occupation to adjust based on the presence or absence of an ELG central, capturing the observed conformity in the data.

For all tracers we account for velocity dispersion by modifying the line-of-sight (LoS) velocities with two additional parameters:
\begin{equation}
\begin{aligned}
    v_{\mathrm{cen,LoS}}^{b} &= v_{\mathrm{halo,LoS}} + \alpha_{\mathrm{cen}} \delta v (\sigma_{\mathrm{LoS}})\,, \\
    v_{\mathrm{sat,LoS}}^{b} &= v_{\mathrm{halo,LoS}} + \alpha_{\mathrm{sat}} (v_{\mathrm{sat,LoS}}-v_{\mathrm{halo,LoS}})\,. \label{eqn:velobias}
\end{aligned}
\end{equation}
Here, $v_{\mathrm{halo,LoS}}$ is the LoS component of the host halo's velocity and $\delta v (\sigma_{\mathrm{LoS}})$ represents the corresponding velocity dispersion. The term $v_{\mathrm{sat,LoS}}$ denotes the satellite galaxy's LoS velocity prior to applying the velocity bias. The parameters $\alpha_{\mathrm{cen}}$ and $\alpha_{\mathrm{sat}}$ capture the deviations of central and satellite velocities from the host halo's velocity and the local dark matter environment, respectively. For a simulation box, the LoS is defined as a fixed, parallel direction corresponding to the viewing direction of an observer at infinity.

\subsubsection{Constructing HOD-informed priors}
\label{subsubsec:hip_construct}

\input{Tables/hoddist}

\begin{figure}
    \centering
    \includegraphics[width=0.95\linewidth]{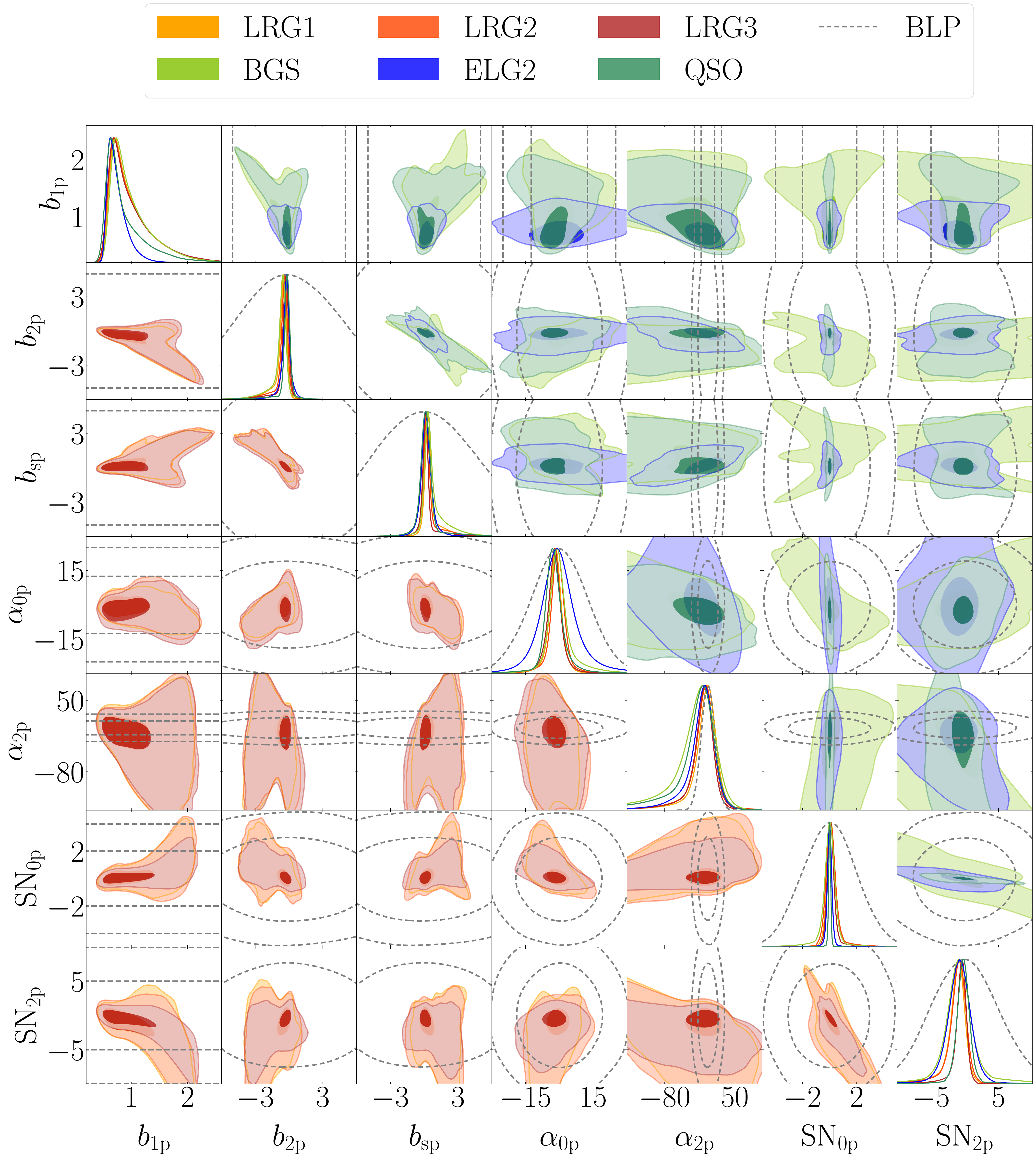}
    \caption{NF-learned HOD-informed priors for all DESI Full-Shape tracers in the physical basis using the EPT velocileptors framework. The lower triangle shows the priors for LRG1, LRG2, and LRG3, while the upper triangle presents those for BGS, ELG, and QSO. The dashed gray contours (and boundary lines) indicate priors used in the baseline study (BLP) for comparison.}
    \label{fig:prior_all}
\end{figure}

Using the HOD prescriptions described above, we generate real-space mock galaxy catalogs for each DESI tracer within the 32 \textsc{AbacusSummit} simulation \cite{Garrison2016ImprovingSimulations,Garrison2018TheSimulations,Garrison2019AABACUS,Garrison2021TheCode,Bose2022ConstructingABACUSSUMMIT} boxes also considered in \cite{Zhang2024HOD-informedLSS}. These boxes collectively span a 7-dimensional cosmological parameter space $\{ \omega_b, \omega_c, h, \ln 10^{10} A_s, n_s, w_0, w_a \}$, enabling us to sample variations in the cosmological parameters of interest (see \cref{app:cosmo_dp} for a detailed description of the simulation boxes used). To convert the catalogs into redshift space, we apply redshift-space distortions (RSD) to the galaxy positions as follows:
\begin{equation}
\mathbf{r}_{\mathrm{redshift}} = \mathbf{r}_{\mathrm{real}} + \frac{\left( \mathbf{v}_{\mathrm{pec}} \cdot \hat{\mathbf{r}}_{\mathrm{LoS}} \right)(1+z)}{H(z)}\,\hat{\mathbf{r}}_{\mathrm{LoS}},
\label{eq:rsd}
\end{equation}
where $\mathbf{r}_{\mathrm{real}}$ denotes the real-space position of the galaxy, $\mathbf{v}_{\mathrm{pec}}$ is the peculiar velocity, $\hat{\mathbf{r}}_{\mathrm{LoS}}$ is the unit vector along the line of sight, and $H(z)$ is the Hubble parameter at redshift $z$. The $(1+z)$ factor converts the displacement into comoving units. Following the pipeline outlined in \cite{Zhang2024HOD-informedLSS}, within each simulation box, and for each tracer, we sample 10,000 distinct HOD parameter sets drawn from a distribution tailored to the specific tracer, resulting in a total of 320,000 measurements per tracer. While the HOD models for each tracer are motivated by the corresponding HOD analyses of DESI EDR data \cite{Yuan2024TheABACUSSUMMIT, Rocher2023TheSimulations, Smith2024GeneratingSurvey}, within the confines of each model, we explore a broad range of HOD parameters to construct our HOD-informed priors for mitigating projection effects. The HOD parameter distributions and the corresponding simulation redshifts used for these measurements are summarized in \cref{tab:hoddist}. These HOD parameter distributions are tuned to have the same peak around the HOD parameter values learned from the corresponding small-scale HOD analysis of that tracer using EDR, but with a much broader range compared to the statistical error, in order to conservatively remove unphysical regions of the nuisance parameter space without imposing small-scale clustering information. As demonstrated in \cite{Zhang2024HOD-informedLSS}, although the specific HOD parameter distributions affect the detailed shape of the resulting priors, they have little impact on the derived cosmological constraints using those priors.

Several studies have indicated that assembly bias \cite{Wechsler2002ConcentrationsHistories,Gao2005TheClustering, Croton2007HaloClustering, Gao2007AssemblyHaloes, Lin2016OnPopulations,Pujol2017WhatDensity, Artale2018TheSimulations, Zehavi2018TheHalos, Hadzhiyska2020LimitationsBeyond, Xu2021DissectingBias, Xu2021PredictingLearning, Delgado2022ModellingLearning, Salcedo2022ElucidatingSDSS, Yuan2022IllustratingILLUSTRISTNG, Wang2022EvidenceStatistics, Yuan2023UnravelingData}—the dependence of halo clustering on secondary properties beyond halo mass—can modify the shape of the HOD-informed prior \cite{2024arXiv241008998A, 2024arXiv241206886S}. However, two mitigating factors come into play. First, the shape changes introduced by a decorated model can, to some extent, be degenerate with adjustments to the HOD parameter distribution in a simpler model \cite{Zhang2024HOD-informedLSS}. Second, the HOD-informed priors derived from both decorated and simpler models largely overlap \cite{2024arXiv241008998A}. Consequently, the impact on the resulting cosmological constraints is expected to be minimal, although a systematic study of these effects will be pursued in future work.

For each mock galaxy catalog, we employ the \texttt{AbacusHOD} code \cite{Hadzhiyska2022COMPASO:Overdensities, Yuan2022ABACUSHOD:Data} to rapidly generate power spectrum multipole measurements. We then fit these to determine the EFT-nuisance parameters of the model described in \cref{subsubsec:fs_model}, for the likelihood maximum, keeping the cosmological parameters fixed to those of the simulation box, i.e. the true values. In our analysis, we focus on fitting the monopole and quadrupole over the same wavenumber range as used in baseline analysis.  Given the large number of measurements, we employ \texttt{Effort.jl}\footnote{\url{https://github.com/CosmologicalEmulators/Effort.jl}} \cite{Bonici2025Effort:Universe}—an emulator capable of reproducing any PT-based theory code—to emulate the EPT \texttt{velocileptors} implementation and to dramatically speed up our analysis. We calculate likelihoods using code written in the probabilistic programming language \texttt{Turing.jl}\footnote{\url{https://github.com/TuringLang/Turing.jl}} \cite{Ge2018Turing:Inference,Fjelde2025Turing.jl:Language}, which allows auto-differentiation enabling us to compute best-fits using the fast L-BFGS minimization algorithm~\cite{Liu1989OnOptimization}; for the robustness checks on the goodness of fits, we refer to Section 3.2 of \cite{Zhang2024HOD-informedLSS}. We search for the best-fit values of the physical basis parameters described in \cref{subsubsec:fs_model}, with $b_{\rm 3p}$, $\alpha_{\rm 4p}$, and ${\rm SN}_{\rm 4p}$ fixed to zero. This search is performed using a Gaussian analytic covariance matrix from the \texttt{CovaPT}\footnote{\url{https://github.com/JayWadekar/CovaPT}} code. We find that the distribution of best-fit values is largely insensitive to the choice of covariance matrix used in the fitting procedure (see Appendix A of \cite{Zhang2024HOD-informedLSS} for further details). It is also worth noting that although we adopt the physical basis implemented in \texttt{velocileptors} for this work, the best-fit values can be readily converted \cite{Holm2023BayesianData} to other parameterizations used by alternative theoretical codes.

Finally, we stack the best-fit parameters \cite{Sullivan2021AnGalaxies} from all 32 simulation boxes for each tracer (at their corresponding redshifts) and employ a Normalizing Flow (NF) model \cite{JimenezRezende2015VariationalFlows} to learn the distribution of these best-fit values.
This approach largely eliminates cosmological dependencies (see \cref{app:cosmo_dp} for further discussion), enabling us to construct a HOD-informed prior that robustly accommodates arbitrary nuisance parameter sets within the explored range.
We implement our NF model using the \texttt{nflows}\footnote{\url{https://github.com/bayesiains/nflows}} \cite{Durkan2019NeuralFlows} library. In particular, our model belongs to the family of Masked Autoregressive Flows (MAF), employing 7 features and 10 layers of Masked Affine Autoregressive Transforms, with reverse permutations included to enhance expressiveness. For further details on the NF model setup and its robustness, see \cite{Zhang2024HOD-informedLSS}.

The resulting NF-learned HOD-informed priors for all DESI tracers are shown in \cref{fig:prior_all}. In the lower triangle, the priors for LRG1, LRG2, and LRG3 are displayed, while the upper triangle presents those for BGS, ELG, and QSO. The dashed gray contours represent the priors used in the baseline study (see Table 1 in \cite{DESICollaboration2024DESIMeasurements}) for comparison. For the LRG samples, which employ the same HOD model and parameter distribution, the NF–learned priors are nearly identical across the three redshift bins. This uniformity results from our re-parameterization into the physical basis, where the inclusion of the $\sigma_{8}(z)$ factor largely removes the redshift dependence of the priors. The differences in the prior widths for the SN terms arise from variations in the shot noise at the corresponding redshifts used for rescaling 
Although BGS and LRG share the same HOD model, they exhibit slightly different HOD parameter distributions and scaling parameters, resulting in a similar overall shape but with differing widths. In contrast, for ELGs, the different HOD model and parameter distribution yield a higher fraction of low-mass halos hosting galaxies, which reduces the overall linear galaxy bias. Consequently, we observe a slight peak shift in the one-dimensional distribution and a more concentrated $b_{\rm 1p}$ distribution. For QSOs, a lower bound on $\log M_{\rm cut}$ also leads to a somewhat more concentrated $b_{\rm 1p}$ distribution. It is also noteworthy that the narrow ${\rm SN}_{\rm 0p}$ prior for QSOs is a consequence of the high shot noise value used for rescaling. Additionally, while most HIP-inferred distributions are narrower than those obtained with the baseline prior, the distribution for $\alpha_{\rm 2p}$ is noticeably broader. One possible explanation is that $\alpha_{\rm 2p}$ is closely tied to the modeling of small-scale velocity dispersion; exploring a wider range of HOD parameters allows us to absorb more of the Fingers-of-God (FoG) anisotropy—a contribution that is often underestimated when using the baseline prior, as suggested by \cite{2025arXiv250110587B}. Validation of the accuracy of the cosmological constraints obtained using the learned HOD-informed priors, as well as the demonstration of its ability to remove projection effects using noiseless synthetic datasets, is provided in \cref{app:validation}.

\section{Likelihoods and cosmological inference}
\label{sec:likelihoods_inference}

We now introduce the likelihoods used in our analysis when fitting to the data and outline our cosmological inference approach. We briefly describe the data sets involved, ranging from DESI Full-Shape measurements to external probes, and then discuss the key components of our inference pipeline, including the Boltzmann solver, sampling methods, and tools for parameter estimation and visualization.

\subsection{Likelihoods}
\label{subsec:likelihoods}

\input{Tables/likelihoods}

Descriptions of all likelihoods used in our analysis are summarized in \cref{tab:likelihoods}, along with the corresponding references. We begin by introducing the DESI likelihood, starting with DESI-FS, which uses the data and modeling framework presented in \cref{subsubsec:fs_model} (see also \cite{DESICollaboration2024DESIQuasarsb} for full details). In this work, we adopt the HOD-informed prior as the default choice for the FS model nuisance parameters. For comparison purposes, we label the prior used by baseline analysis \cite{DESICollaboration2024DESIQuasarsb,DESICollaboration2024DESIMeasurements} as (BLP) and the HOD-informed prior as (HIP) when required. The DESI-BAO likelihood corresponds to the post-reconstruction BAO-only measurements described in \cite{Adame2025DESIOscillations}, including all six DESI tracer samples as well as the Ly$\alpha$ BAO likelihood \cite{DESICollaboration2024DESIQuasars,Adame2025DESIForest}. The combined DESI likelihood (DESI) includes both the Full-Shape (FS) and BAO information, with correlations between the power spectrum and post-reconstruction BAO measurements consistently accounted for via the full mock-based covariance matrix.

In addition to the likelihoods from DESI, we combine our results with multiple external data sets. This includes CMB likelihoods CMB-nl, CMB and CMB-PR4. The CMB-nl likelihood is based on the \emph{Planck} 2018 PR3 release \cite{Collaboration2020PlanckParameters}, which includes temperature (TT) and polarization (EE) auto-spectra, plus their cross-spectra (TE), using the \texttt{simall}, \texttt{Commander}, and \texttt{plik} likelihoods. The CMB likelihood augments CMB-nl with the latest Planck PR4 (\texttt{NPIPE}) CMB lensing data \cite{Carron2022CMBMaps} and Atacama Cosmology Telescope (ACT) lensing \cite{Madhavacheril2024TheParameters,Qu2024TheGrowth,MacCrann2024TheAnalysis}. Finally, CMB-PR4 is a \emph{Planck} PR4–based likelihood \cite{Tristram2021PlanckRatio,Tristram2024CosmologicalPR4} (including \texttt{LoLLiPoP} for low-$\ell$ and \texttt{HiLLiPoP} for high-$\ell$) that is also combined with lensing. 

We further include Type Ia supernova (SN Ia) data, which serve as standardizable candles for measuring the cosmic expansion history. Specifically, we employ type Ia supernova (SN Ia) datasets from PantheonPlus \cite{Brout2022TheConstraints,Scolnic2022TheRelease}, Union3 \cite{Rubin2023UnionFramework}, and Dark Energy Survey Year 5 supernova analysis \cite{Collaboration2024TheSet}. 
Finally, we impose Big Bang Nucleosynthesis (BBN) \cite{Schoneberg2024TheUpdate} priors on $\Omega_b h^2$ and a loose prior on the spectral index $n_s$ (denoted $n_{\mathrm{s10}}$), which is set to be 10 times broader than the \emph{Planck}-derived posterior width \cite{Collaboration2020PlanckParameters}. These BBN and $n_{\mathrm{s10}}$ priors are included only in scenarios that do not incorporate a CMB likelihood.

\subsection{Cosmological inference}
\label{subsec:cosmo_inference}

Our cosmological inference pipeline largely follows \cite{DESICollaboration2024DESIMeasurements}. We use \texttt{Cobaya} \cite{Torrado2019Cobaya:Cosmology,Torrado2021Cobaya:Models} as the primary framework to explore parameter space, with \texttt{CAMB} \cite{Lewis2000EfficientModels,Howlett2012CMBCosmology} serving as the Boltzmann code for generating linear matter power spectra as a function of cosmological parameters. The Bayesian analysis is performed using a Metropolis–Hastings Markov Chain Monte Carlo (MCMC) sampler \cite{Lewis2002CosmologicalApproach, Lewis2013EfficientParameters}. We then employ \texttt{GetDist} \cite{Lewis2019GetDist:Samples} to extract parameter constraints from the posterior distributions and \texttt{iminuit} \cite{James1975MinuitCorrelations, iminuit} to obtain the maximum a posteriori (MAP) estimates. We use \texttt{velocileptors} to compute the theoretical model for cosmological results, in line with the baseline analysis \cite{DESICollaboration2024DESIQuasarsb, DESICollaboration2024DESIMeasurements}, rather than employing \texttt{effort.jl}.

The convergence criterion for MCMC sampling is that the Gelman-Rubin statistic \cite{1992StaSc...7..457G} satisfies $R-1 < 0.01$ for all cosmological parameters. We utilize 4 walkers, discarding the first 30\% of iterations as burn-in. Unless otherwise noted, the cosmological parameters and priors used here match those in Table~1 of \cite{DESICollaboration2024DESIMeasurements}, while the nuisance parameters in the DESI-FS and DESI likelihood default to the HOD-informed priors.

\section{Results}
\label{sec:results}

In this study, we focus on two dark energy models, $\Lambda\mathrm{CDM}$ and $w_0w_a\mathrm{CDM}$, which are of primary interest given the recent DESI BAO results \cite{Adame2025DESIOscillations,2025arXiv250314738D}. All cosmological constraints obtained from our selected model and dataset combinations are summarized in \cref{tab:results}. 

\subsection{\texorpdfstring{$\Lambda {\rm CDM}$ model}{Lambda CDM model}}
\label{subsec:results_lcdm}

\begin{figure}
    \centering
    \includegraphics[width=0.95\linewidth]{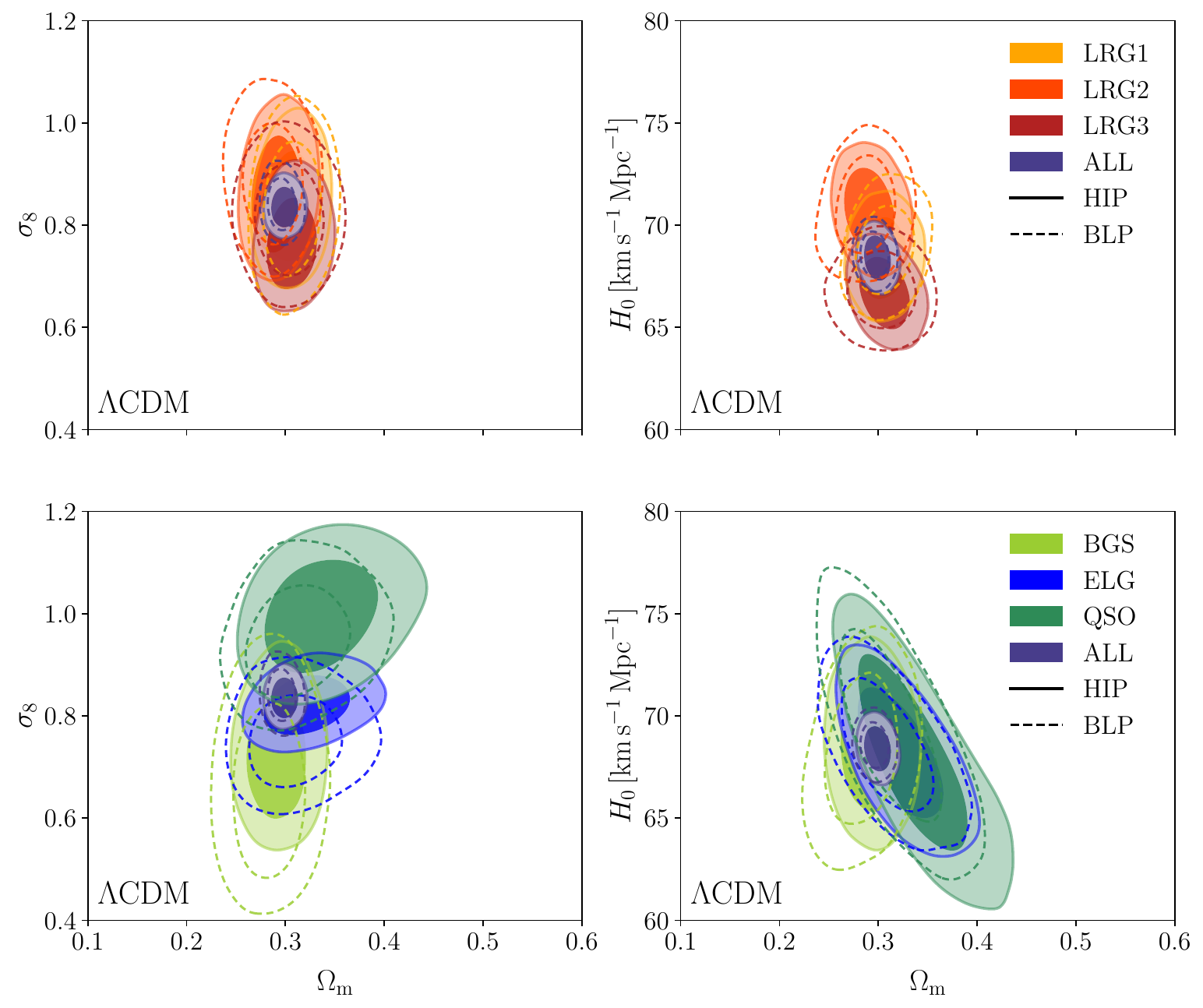}
    \caption{68\% and 95\% credible intervals for $\Omega_{\rm m}$ vs.\ $\sigma_8$ (left panels) and $\Omega_{\rm m}$ vs.\ $H_0$ (right panels) in a flat $\Lambda$CDM model, derived for each individual DESI tracer (FS+BAO) combined with BBN and $n_{\rm s10}$ priors. The darkslateblue contours labeled “ALL” correspond to the combination of all DESI tracers. Results are shown for two sets of nuisance-parameter priors: the HOD-informed prior (HIP, solid filled) and the baseline prior (BLP, dashed unfilled). Tracers are arranged in two rows for enhanced visualization.}
    \label{fig:lcdm_tracer}
\end{figure}

\begin{figure}
    \centering
    \includegraphics[width=0.95\linewidth]{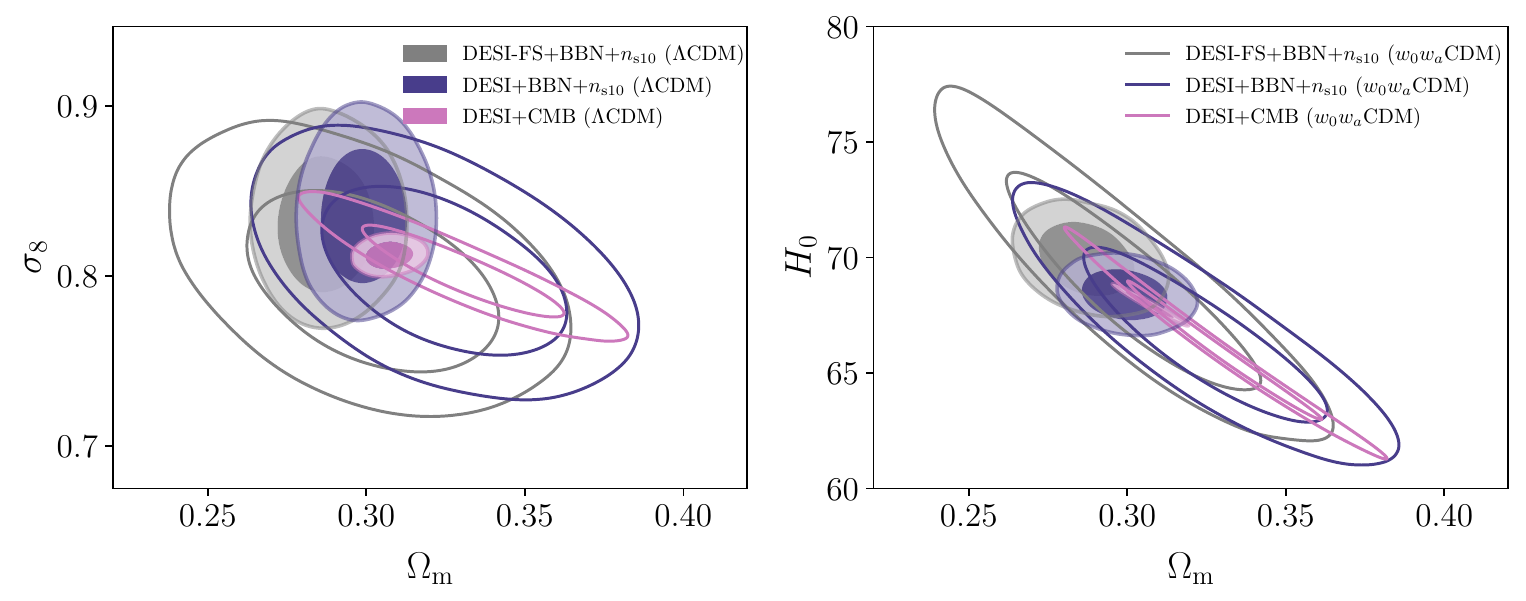}
    \caption{68\% and 95\% credible intervals for $\Omega_{\rm m}$ vs.\ $\sigma_8$ (left) and $\Omega_{\rm m}$ vs.\ $H_0$ (right), all use HOD-informed priors. Filled contours correspond to the $\Lambda\mathrm{CDM}$ model, while unfilled contours represent the $w_0w_a\mathrm{CDM}$ extension. The color scheme distinguishes different data combinations—DESI-FS+BBN+$n_{\rm s10}$, DESI+BBN+$n_{\rm s10}$, and DESI+CMB—as indicated in the legend.
}
    \label{fig:Om_sigma8_H0}
\end{figure}

We begin by discussing our results within a flat $\Lambda\mathrm{CDM}$ framework. \cref{fig:lcdm_tracer} compares the constraints from each DESI tracer (FS+BAO combined with BBN and $n_{\rm s10}$ priors) under two choices for the FS nuisance parameter priors: the HOD-informed prior (HIP) and the baseline prior (BLP). Each panel displays the two-dimensional posterior contours for $\Omega_{\rm m}$ versus $\sigma_8$ or $H_0$. Overall, we observe strong consistency between HIP and BLP, and the introduction of HOD-informed priors does not significantly shift the constraints.

For LRGs, the HIP shifts $\sigma_8$ to slightly lower values compared to the BLP, while the mean values of $\Omega_{\rm m}$ and $H_0$ remain largely unchanged. For BGS, ELG, and QSO, the HIP tends to shift $\Omega_{\rm m}$ and $\sigma_8$ to marginally higher values, with $H_0$ staying roughly constant. Additionally, the reduction in contour size is more pronounced for tracers with lower signal-to-noise ratios—for example, the decrease in contour size is most evident for LRG3 relative to LRG1 and LRG2. The dark slate blue contours represent the combined results from all tracers. Compared to the BLP, the HIP only yields small shifts in the mean (approximately $+0.33\sigma$ for $\Omega_{\rm m}$, $-0.17\sigma$ for $\sigma_8$, and $-0.23\sigma$ for $H_0$) and tighter constraints by 4\%, 23\%, and 2\% for $\Omega_{\rm m}$, $\sigma_8$, and $H_0$, respectively. These findings are consistent with the expectation that HIP effectively remove unphysical regions of parameter space, thereby enhancing the overall cosmological constraints without significantly altering the central values for models, such as flat $\Lambda\mathrm{CDM}$, where the parameters are well constrained. It is also worth noting that the two parameters showing less improvement—$H_0$ and $\Omega_{\rm m}$—are primarily constrained by BAO measurements, which are less sensitive to the broadband shifts that HOD-informed priors more effectively capture.

Focusing on the results using HIP, a zoomed-in view of the combined DESI contour (darkslateblue filled) is shown in \cref{fig:Om_sigma8_H0}, along with a comparison to the DESI-FS case (gray filled). We obtain the following constraints on these three parameters:
\begin{equation}
\left.
 \begin{aligned}
\Omega_{\rm m} &= 0.2874^{+0.0094}_{-0.010} \\
\sigma_8 &= 0.832^{+0.024}_{-0.028} \\
H_0      &= (69.90\,\pm\,1.0)\,{\rm km\,s^{-1} Mpc^{-1}}
 \end{aligned}
\ \right\}
\ \mbox{\text{\leftparbox{3.5cm}{\text{DESI-FS+BBN+$n_{\rm s10}$,}}}}
\label{eqn:lcdm_desi_fs}
\end{equation}

\begin{equation}
\left.
 \begin{aligned}
\Omega_{\rm m} &= 0.2909\,\pm\,0.0090 \\
\sigma_8 &= 0.836^{+0.024}_{-0.027} \\
H_0      &= (68.40\,\pm\,0.73)\,{\rm km\,s^{-1} Mpc^{-1}}
 \end{aligned}
\ \right\}
\ \mbox{\text{\leftparbox{3.5cm}{\text{DESI+BBN+$n_{\rm s10}$.}}}}
\label{eqn:lcdm_desi_fsbao}
\end{equation}
In accordance with the notation in \cref{tab:likelihoods}, “DESI-FS” denotes the Full-Shape only likelihood, whereas “DESI” refers to the combined FS and BAO likelihood. In both cases, BBN and $n_{\rm s10}$ priors are applied.

We observe a clear shift in the central values of $\Omega_{\rm m}$ and $H_0$ when reconstructed BAO information is added to the FS likelihood, consistent with the shift reported in \cite{DESICollaboration2024DESIQuasarsb}. Specifically, the shift corresponds to approximately $0.37\sigma$ for $\Omega_{\rm m}$ and $1.5\sigma$ for $H_0$, while $\sigma_8$ remains unchanged, consistent with BAO being a purely geometric probe that is insensitive to $\sigma_8$. The combined DESI likelihood improves the constraints on $\Omega_{\rm m}$ and $H_0$ by roughly 4\% and 27\%, respectively, compare to DESI-FS. 

We consider the parameter combination $S_8 = \sigma_8\left({\Omega_{\rm m}}/{0.3}\right)^{0.5}$, which is known to be tightly constrained by weak lensing analyses. For DESI combined with BBN and $n_{\rm s10}$ priors, we obtain
\begin{equation}
\left.
 S_8 = 0.835^{+0.027}_{-0.030} \quad\mbox{\text{\leftparbox{3.5cm}{(DESI+BBN+$n_{\rm s10}$).}}}
\right.
\end{equation}
This result is 17\% tighter compared to the constraint obtained using the baseline prior, while the mean value remains largely unchanged. We refer the reader to \cite{DESICollaboration2024DESIMeasurements} for a detailed discussion of the $S_8$ constraints.

We now turn to the combination of DESI with CMB data. In particular, when DESI is combined with the CMB likelihood, we obtain:
\begin{equation}
\left.
 \begin{aligned}
\Omega_{\rm m} &= 0.3072\,\pm\,0.0049 \\
\sigma_8 &= 0.8123\,\pm\,0.0052 \\
H_0      &= (67.95\,\pm\,0.37)\,{\rm km\,s^{-1} Mpc^{-1}}
 \end{aligned}
\ \right\}
\ \mbox{\text{\leftparbox{3.8cm}{\text{DESI+CMB.}}}}
\label{eq:desi_cmb_lcdm}
\end{equation}
The corresponding contours are shown in \cref{fig:Om_sigma8_H0} (pink filled). We find that DESI and the CMB data sets are highly consistent, with no significant shift in the central values beyond expected statistical fluctuations. We also consider two alternative CMB likelihoods: CMB-nl, which omits the lensing reconstruction, and CMB-PR4, which utilizes the \textit{Planck} PR4 data release. Focusing on $\sigma_8$, the DESI+CMB-nl combination yields a $\sim34\%$ larger uncertainty compared to DESI+CMB, whereas DESI+CMB-PR4 achieves a $\sim13\%$ reduction. Detailed constraints are provided in \cref{tab:results}.

\subsection{\texorpdfstring{$w_0 w_a {\rm CDM}$ model}{w0wa CDM model}}

\label{subsec:results_wcdm}
\begin{figure}
    \centering
    \includegraphics[width=0.95\linewidth]{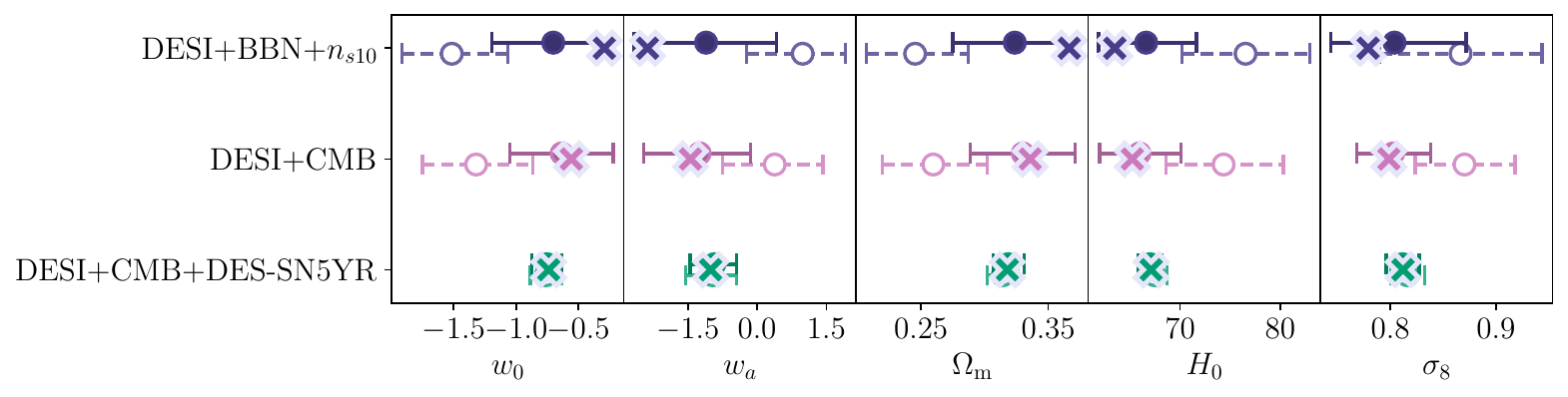}
    \caption{Illustration of projection effects in a $w_0w_a$CDM analysis, comparing three data combinations: DESI+BBN+$n_{\rm s10}$, DESI+CMB, and DESI+CMB+DES-SN5YR. For each combination, we plot the mean and 95\% one-dimensional credible intervals for both the baseline prior (BLP) and the HOD-informed prior (HIP). To improve visibility, the BLP (unfilled markers with dashed error bars) and HIP (filled markers with solid error bars) points are slightly offset along the vertical axis. Cross markers denote the maximum a posteriori (MAP) estimate under BLP.}
    \label{fig:projection_1d}
\end{figure}

\begin{figure}
    \centering
    \includegraphics[width=0.95\linewidth]{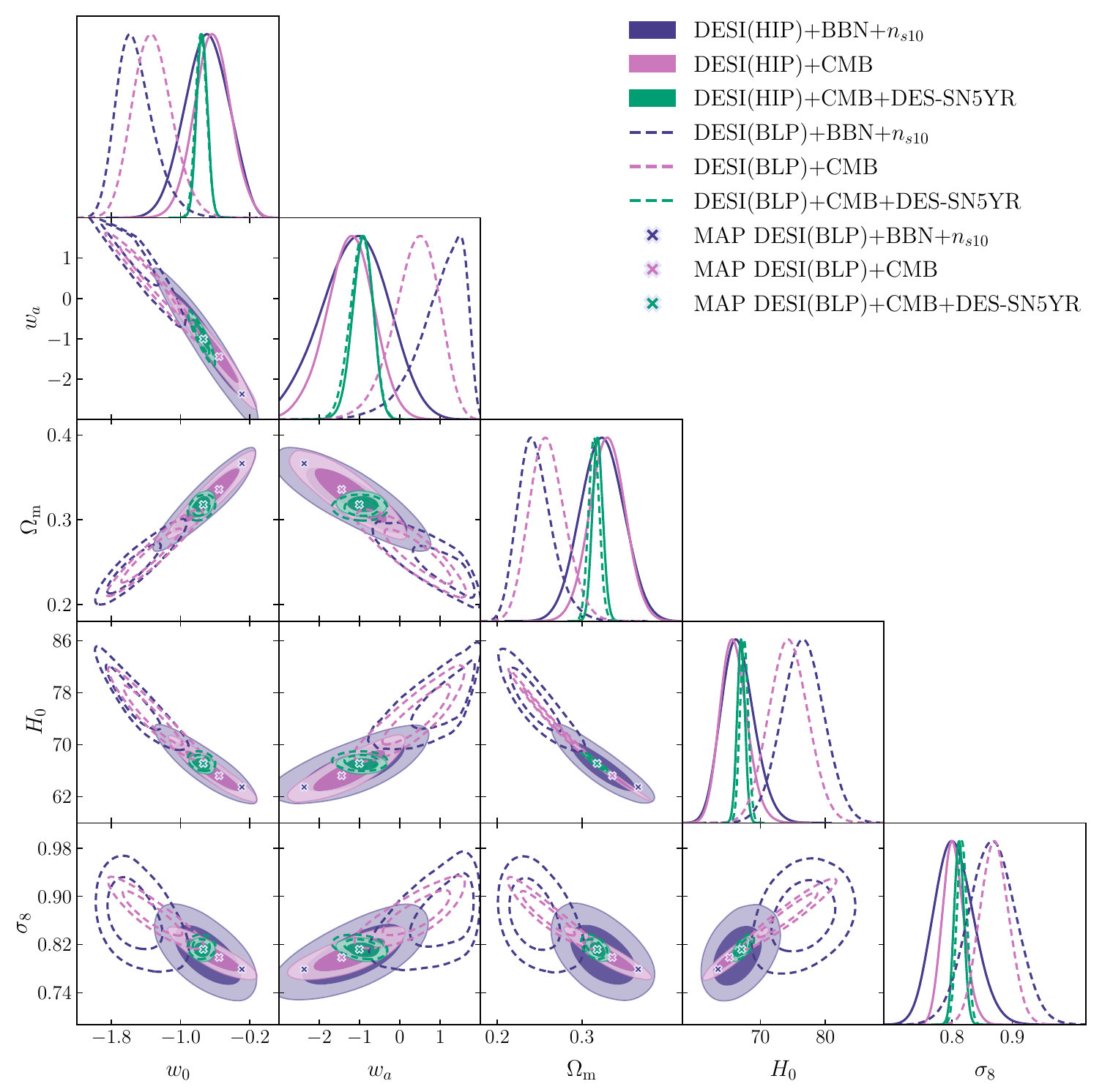}
    \caption{68\% and 95\% credible intervals for a range of data combinations under the $w_0w_a$CDM model. Filled contours (solid lines) are derived using HOD-informed priors (HIP), whereas unfilled contours (dashed lines) correspond to the baseline prior (BLP). Markers indicate the maximum a posteriori (MAP) values obtained with the BLP.}
    \label{fig:projection}
\end{figure}

We study the time-varying dark energy equation of state in the following parameterization, 
\begin{equation}
    w(a) =w_0+w_a(1-a),
    \label{eqn:cpl}
\end{equation}
allowing extra freedom in the expansion history. Before presenting our cosmological results in the $w_0w_a$CDM model, we revisit the severe projection effects observed in the baseline analysis that employed Gaussian priors for most nuisance parameters (Appendix~A of \cite{DESICollaboration2024DESIMeasurements}). In that approach, the extensive nuisance parameter volume introduced additional degeneracy directions, causing the maximum a posteriori (MAP) estimates to fall outside the 95\% Bayesian credible intervals for the FS, FS+BAO, and FS+BAO+CMB likelihoods. This indicates that the baseline priors on nuisance parameters allowed large regions of parameter space to persist, which were instead discarded by the HIP, ultimately affecting the constraints on cosmological parameters.

Validation tests using noiseless synthetic data (see \cref{app:validation}) indicated that applying HIP to the nuisance parameters effectively alleviates projection effects and accurately recovers the underlying cosmology. In the survey data, \cref{fig:projection_1d} and \cref{fig:projection} display the maximum a posteriori (MAP) estimates along with the 1D and 2D Bayesian credible intervals for DESI, DESI+CMB, and DESI+CMB+DES-SN5YR under both HIP and BLP. 
However, for these comparisons we adopt the MAP values obtained with the BLP rather than those from the HIP. This is because the HIP exhibits a more complex geometry, and we are unable to perform analytical marginalization for the counterterms and shot noise components, resulting in a higher effective dimensionality of the nuisance parameter space that makes maximization of the posterior significantly more challenging. Due to these numerical difficulties, we do not report the HIP MAP values; instead, we use the BLP MAP as a proxy, which is justified by the observation that the central values of the priors remain largely unchanged (see \cref{fig:prior_all}). In future work, we plan to develop a much faster inference pipeline—incorporating an emulator for theory computation and optimized minimization algorithms—or to apply additional smoothing to the prior to enhance its stability, thereby improving the robustness of MAP estimation.

\cref{fig:projection_1d} displays the MAP estimates (obtained using the BLP) together with the 95\% Bayesian intervals computed under both HIP and BLP for DESI, DESI+CMB, and DESI+CMB+DES-SN5YR. In nearly all cases, the MAP estimates fall well within the 95\% credible intervals, confirming the robustness of our HIP-based approach, although for the DESI+BBN+$n_{\rm s10}$ case some MAP values lie near the interval boundary, indicating a potential residual projection effect (see \cref{app:cosmo_dp} for more discussion). Nevertheless, this residual effect is much less pronounced than that observed under the BLP. \cref{fig:projection} presents the two-dimensional posterior contours for the same likelihood combinations, comparing HIP (solid filled contours) with BLP (dashed unfilled contours). For both DESI and DESI+CMB, the HIP contours are noticeably shifted toward the MAP, reflecting the effect of covering only a physical region of the nuisance parameter space. This improved alignment between the MAP and the Bayesian credible intervals clearly demonstrates that our HIP approach effectively mitigates the projection issues observed with the BLP.

Before presenting constraints on $(w_0, w_a)$, we revisit \cref{fig:Om_sigma8_H0}, which displays the posterior contours for $\Omega_{\rm m}$-$\sigma_8$ and $\Omega_{\rm m}$-$H_0$ under both the flat $\Lambda$CDM and $w_0w_a$CDM models. As expected, the inclusion of two additional dark energy parameters in the $w_0w_a$CDM model enlarges the contours due to the increased parameter space. However, the shifts in the central values of $\Omega_{\rm m}$ and $H_0$ between the DESI-FS and DESI (FS+BAO) cases persist, mirroring those observed in the $\Lambda$CDM analysis.

As for $(w_0, w_a)$, the combination DESI+BBN+$n_{\rm s10}$ yields:
\begin{equation}
\left.
 \begin{aligned}
w_0 &= -0.70\,\pm\,0.25 \\
w_a &= -1.10\,\pm\,0.79
 \end{aligned}
\ \right\}
\ \mbox{\text{\leftparbox{3.0cm}{DESI+BBN+$n_{\rm s10}$,}}}
\label{eqn:w0wa_desi_bbn}
\end{equation}
and when we combine DESI with CMB data, we obtain:
\begin{equation}
\left.
 \begin{aligned}
w_0 &= -0.63\,\pm\,0.21 \\
w_a &= -1.24^{+0.61}_{-0.55}
 \end{aligned}
\ \right\}
\ \mbox{\text{\leftparbox{3.0cm}{DESI+CMB.}}}
\label{eqn:w0wa_desi_cmb_bbn}
\end{equation}
We also provide results for DESI-FS+BBN+$n_{\rm s10}$, as well as the combination of DESI with CMB-nl and CMB-PR4, in \cref{tab:results}.

\begin{figure}
    \centering
    \includegraphics[width=\linewidth]{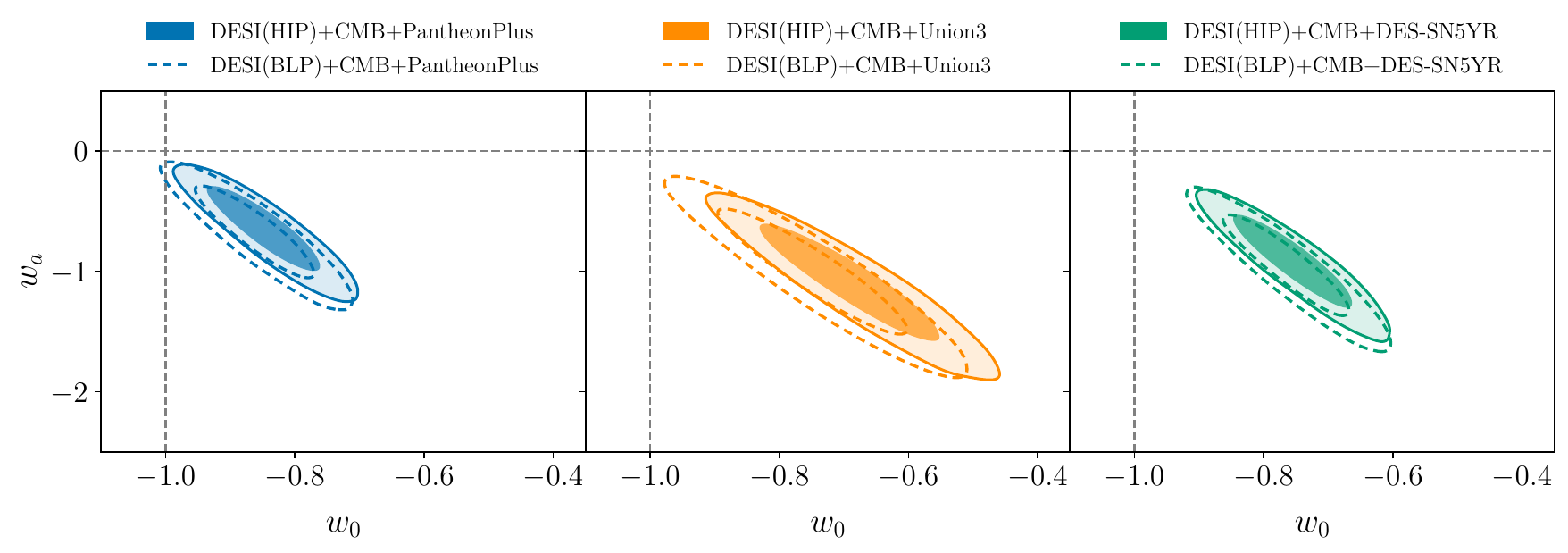}
    \caption{Constraints on $(w_0, w_a)$ from DESI+CMB combined with three different supernova datasets (PantheonPlus, Union3, and DES-SN5YR). Each panel corresponds to one of the SN datasets, with filled contours (solid lines) showing results under the HOD-informed prior (HIP), and unfilled contours (dashed lines) using the baseline prior (BLP).}
    \label{fig:cpl_sne}
\end{figure}

\begin{figure}
    \centering
    \begin{subfigure}{0.48\linewidth}
        \centering
        \includegraphics[width=\linewidth]{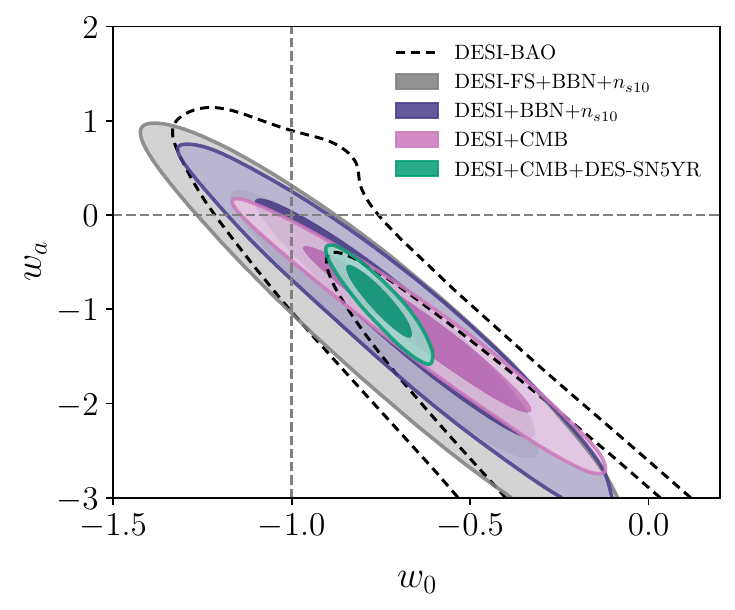}
        \caption{Constraints from multiple DESI data combinations as listed in \cref{tab:results} are shown, including DESI-FS (FS) and DESI (FS+BAO) combined with BBN and $n_{\rm s10}$, as well as a combination that adds CMB and the DES-SN5YR supernova dataset. For clarity, only one SN dataset (DES-SN5YR) is shown. The unfilled black dashed contour corresponds to the DESI BAO–only result for comparison.}
        \label{fig:cpl}
    \end{subfigure}
    \hfill
    \begin{subfigure}{0.48\linewidth}
        \centering
        \includegraphics[width=\linewidth]{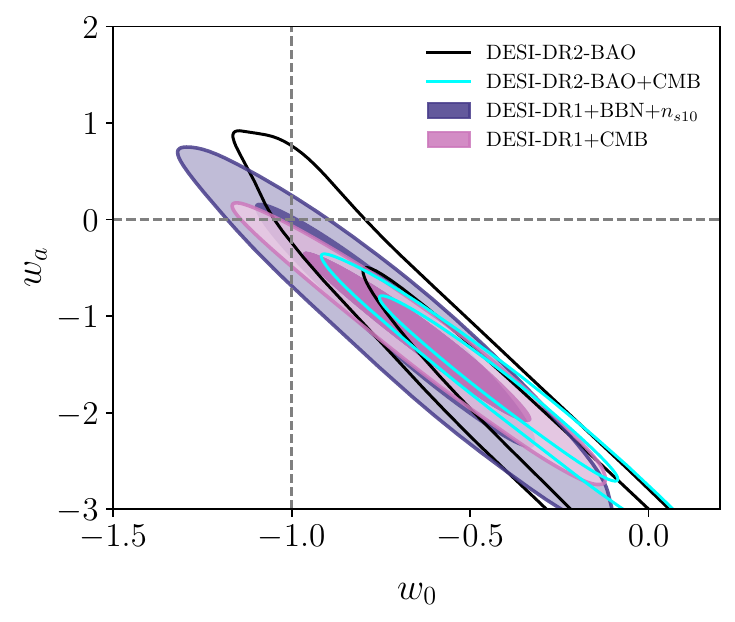}
        \caption{Constraints from different DESI data releases, highlighting a comparison between DR1 FS+BAO and DR2 BAO-only measurements. The black and cyan lines correspond to DESI-DR2-BAO only and DESI-DR2-BAO+CMB, respectively, while the slateblue and pink contours are adopted from the left panel. The legend (DR1/DR2) indicates the specific data release associated with each measurement.}
        \label{fig:cpl_y3bao}
    \end{subfigure}
    \caption{Constraints on $(w_0, w_a)$ from the various data combinations, HOD-informed priors (HIP) are employed for all Full-Shape related likelihoods.}
    \label{fig:cpl_all}
\end{figure}

\cref{fig:cpl_sne} compares DESI+CMB combined with three different SN~Ia datasets: PantheonPlus, Union3, and DES-SN5YR, under the HOD-informed prior (HIP, solid filled contours) versus the baseline prior (BLP, unfilled dashed). Across all SN~Ia samples, the constraints exhibit excellent consistency, with HIP generally providing slightly tighter contours and only minimal shifts in the central values. In fact, the HIP approach yields approximately 5\% tighter constraints on both $w_0$ and $w_a$ compare to BLP.

\cref{fig:cpl} provides a comprehensive view of the $(w_0, w_a)$ plane for a range of data combinations using HIP, with the DESI-BAO–only result (black dashed unfilled contour) serving as a reference. Notably, DESI-FS and DESI (FS+BAO) remain in the same quadrant identified by DESI-BAO, indicating strong consistency within DESI DR1. Each addition—whether it is the Full-Shape information, CMB data, or supernova constraints—shifts or tightens the contours slightly, but all solutions converge to a region with $w_0 > -1$ and $w_a < 0$. This suggests a mild departure from the $\Lambda\mathrm{CDM}$ point $\bigl(-1,\,0\bigr)$ and highlights the possibility of dynamical dark energy. 
\cref{fig:cpl_y3bao} further incorporates the recent DESI DR2 BAO results \cite{2025arXiv250314738D,2025arXiv250314739D} into the comparison, providing a direct contrast between DR2 BAO and the DESI DR1 (FS+BAO) constraints. Overall, the contours from DR1 FS+BAO and DR2 BAO show excellent consistency, demonstrating strong agreement across these data releases. The DR2 BAO constraints, when combined with CMB, yield the tightest contours in this figure, underscoring the improvements in data quality from DR1 to DR2.

\input{Tables/results}

\section{Conclusions}
\label{sec:conclusions}

In this work, we have presented a comprehensive analysis of DESI DR1 Full-Shape and BAO data, supplemented by external datasets from observations of the CMB, SN Ia, and BBN, to constrain cosmological parameters within both the $\Lambda$CDM and $w_0w_a$CDM models. A key addition in our analysis is the introduction of HOD-informed priors (HIP) for the Full-Shape nuisance parameters. We generated $1{,}920{,}000$ power spectrum multipole measurements from a diverse suite of HOD mock catalogs, spanning a wide range of HOD parameters and cosmological models. By fitting the theoretical Full-Shape model to these mock measurements, we extracted best-fit nuisance parameters whose distribution forms the basis of our HOD-informed prior. This approach yields a more physically motivated parameter space by directly linking the nuisance parameters to the underlying galaxy–halo connection and only retaining regions of parameter space where both HOD and EFT models exist. By including mocks spanning a range of cosmological models, we reduce the potential cosmological dependency of our prior.

Within the flat $\Lambda$CDM framework, our analysis yields constraints of $\Omega_{\rm m} = 0.2909\,\pm\,0.0090$, $\sigma_8 = 0.835\,\pm\,0.024$, and $H_0 = 68.4\,\pm\,0.73$ when using the combined FS+BAO DESI likelihood (DESI+BBN+$n_{\rm s10}$). Compared to results obtained using the baseline prior used in \cite{DESICollaboration2024DESIQuasarsb,DESICollaboration2024DESIMeasurements}, the HIP provides approximately 4.2\%, 22.7\%, and 2.0\% tighter constraints on $\Omega_{\rm m}$, $\sigma_8$, and $H_0$, respectively, with non-significant shifts in the central values. 

For the $w_0w_a$CDM model, we find that the application of HIP significantly mitigates the projection effects that plagued previous analyses. Using DESI+BBN+$n_{\rm s10}$ alone, we obtain $w_0 = -0.81\,\pm\,0.07$ and $w_a = -0.70\,\pm\,0.25$; when combined with CMB data, the constraints tighten further to $w_0 = -0.85\,\pm\,0.05$ and $w_a = -0.63\,\pm\,0.18$. In all cases, the maximum a posteriori estimates lie well within the 95\% Bayesian credible intervals, confirming the robustness of our approach.

Overall, our findings demonstrate that the use of HOD-informed priors not only enhances the precision of cosmological constraints but also yields robust dynamical dark energy results from DESI by mitigating projection effects. The mild departures observed in the dark energy parameters—specifically, a trend toward $w_0 > -1$ and $w_a < 0$—offer intriguing hints of dynamical dark energy. We expect that this approach will also be successfully applied to DESI DR2 Full-Shape analyses, which should further refine these constraints and enhance our understanding of dark energy.

\appendix
\begin{figure}
    \centering
    \includegraphics[width=0.95\linewidth]{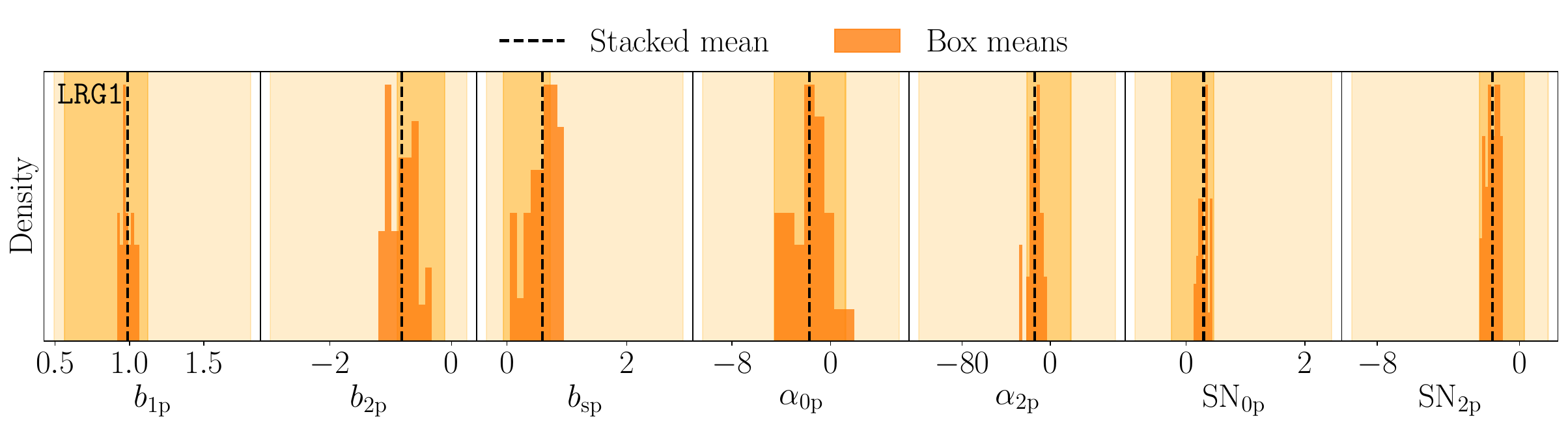}
    \caption{Cosmological dependency of the HOD-informed prior for \texttt{LRG1}. The light orange shaded regions indicate the 1$\sigma$ (lighter shade) and 2$\sigma$ (darker shade) intervals of the HOD-informed prior learned from stacked distribution, while the dark orange histograms show the mean best-fit values from each simulation box. The dashed black line represents the stacked mean.}
    \label{fig:cosmo_dp}
\end{figure}

\begin{figure}
    \centering
    \includegraphics[width=0.95\linewidth]{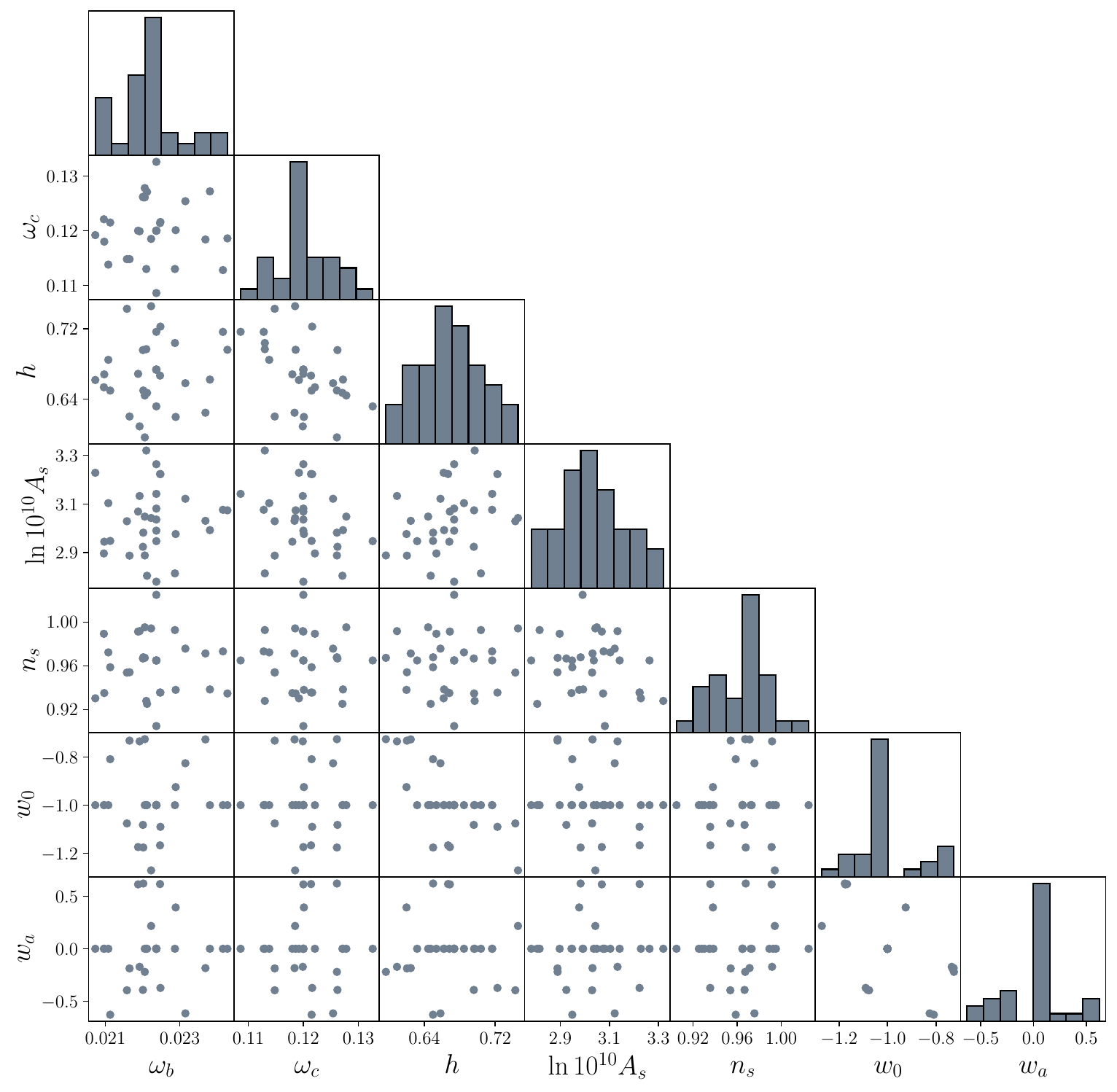}
    \caption{Pairwise scatter plots and marginal histograms illustrating the 7-dimensional cosmological parameter space $\{\omega_b,\, \omega_c,\, h,\, \ln 10^{10} A_s,\, n_s,\, w_0,\, w_a\}$ covered by the 32 simulation boxes used in this work.}
    \label{fig:boxes}
\end{figure}

\begin{figure}
    \centering
    \includegraphics[width=0.95\linewidth]{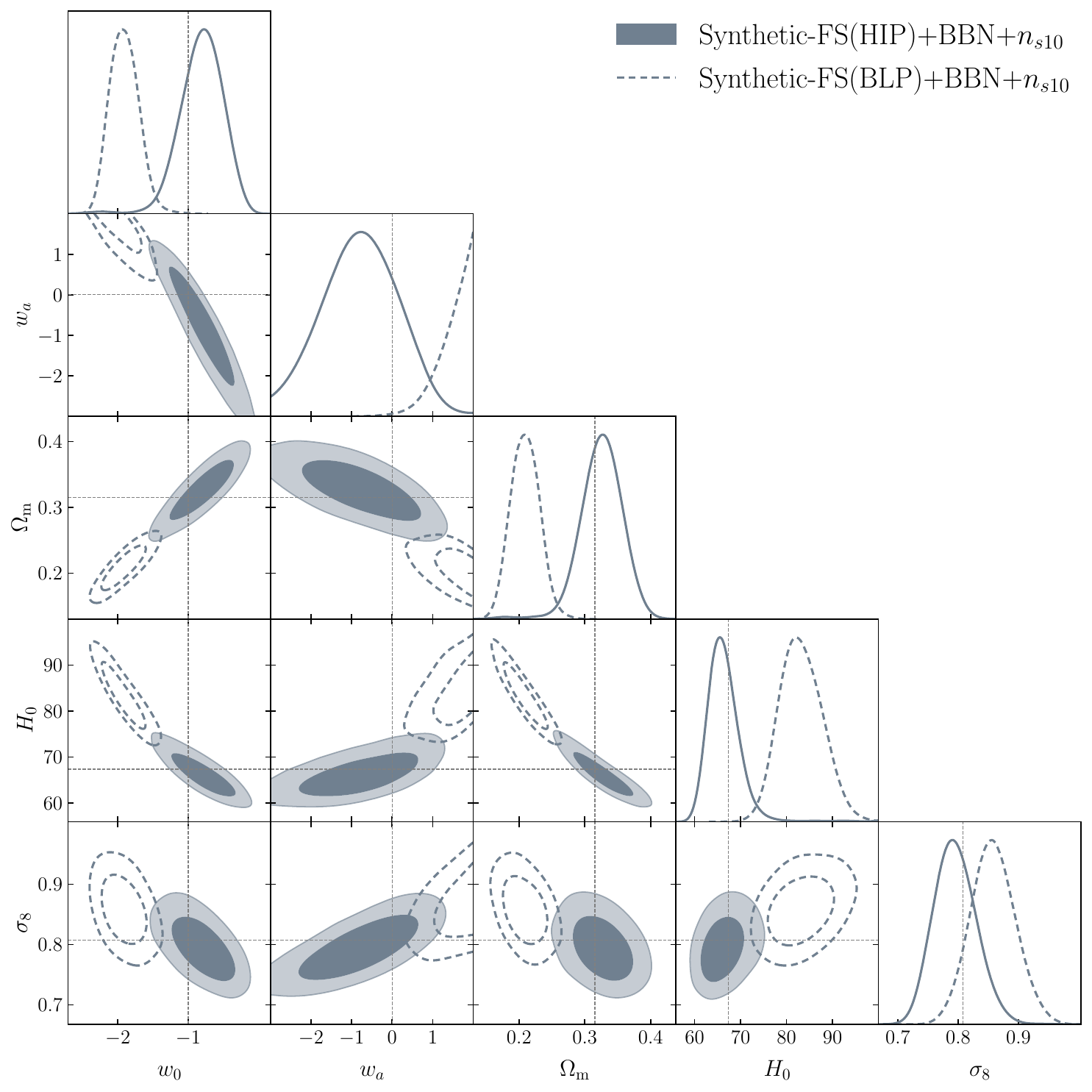}
    \caption{Validation test using synthetic noiseless data generated by the \texttt{velocileptors} code. The filled contours show the posterior obtained with the HOD-informed prior (HIP), while the dashed contours indicate results using the baseline prior (BLP). This test verifies whether HIP can accurately recover the true cosmological parameters in a controlled setup.}
    \label{fig:synthetic}
\end{figure}

\section{Simulation boxes and cosmological dependency}
\label{app:cosmo_dp}

Following our previous work \cite{Zhang2024HOD-informedLSS}, we employed the \texttt{AbacusSummit\_base\_c000\_ph000} simulation box—which corresponds to the Planck 2018 cosmology \cite{Collaboration2020PlanckParameters}—together with boxes ranging from \texttt{AbacusSummit\_base\_c130\_ph000} to \texttt{AbacusSummit\_base\_c160\_ph000}. These additional boxes form an unstructured emulator grid around the \texttt{c000} box and span the seven-dimensional cosmological parameter space $\{\omega_b, \, \omega_c, \, h,\, \ln 10^{10} A_s,\, n_s,\, w_0,\, w_a\}$. These 32 simulation boxes were carefully selected to capture the variations in the cosmological parameters of interest, thereby providing a comprehensive and diverse sampling of the parameter space. \cref{fig:boxes} displays the pairwise projections of these parameters.

By stacking best-fit distributions from 32 distinct cosmologies, we effectively remove cosmological dependencies in the HOD-informed prior within the covered range. \cref{fig:cosmo_dp} overlays the 1$\sigma$ and 2$\sigma$ bands (as shown previously in \cref{fig:prior_all}) for the HOD-informed prior of \texttt{LRG1}—constructed from the stacked distribution—with histograms of the mean best-fit values from 10,000 measurements in each individual box. The variations in the parameter means due to differing cosmologies lie predominantly within the 1$\sigma$ band, indicating that the width of the HIP learned from the stacked distribution is sufficient to encompass the minor differences introduced by changes in cosmology. We present such a figure for \texttt{LRG1} only for simplicity; similar behavior is observed for the other tracers.

However, given that our exploration of the $w_0$ and $w_a$ space is limited, if the data were to favor $w_0$ and $w_a$ values far from $(-1,0)$, it is possible that the corresponding nuisance parameter values would fall outside the current 1$\sigma$ or 2$\sigma$ bands of our HIP. This appears to be the case for the FS+BAO analysis shown in \cref{fig:projection_1d} and \cref{fig:projection}, where the MAP is located in a $w_0$-$w_a$ region not covered by the emulator grid defined by the 32 cosmologies. Consequently, the current version of the HIP may still exhibit residual prior weight effects due to this potential shift in the nuisance parameter space. This issue could be addressed by incorporating additional simulations in the data-preferred $w_0$-$w_a$ quadrant; however, the current number of simulation boxes available in that region is limited.

\section{Validation using synthetic dataset}
\label{app:validation}

We validate our HOD-informed prior (HIP) by performing a synthetic test using the same noiseless dataset adopted in \cite{DESICollaboration2024DESIQuasarsb}. This dataset is generated with the \texttt{velocileptors} model in the physical basis described in \cref{subsec:fs}, where the underlying true cosmology is set to DESI fiducial. The nuisance parameters correspond to the best-fit values derived from the observed FS clustering, ensuring that the synthetic data accurately reflects realistic galaxy clustering signals.

We then fit this noiseless dataset under a $w_0w_a$CDM model to examine whether HIP can correctly recover the true cosmology. As shown in \cref{fig:synthetic}, the posteriors derived with HIP closely encapsulate the input cosmology, exhibiting minimal projection effects compared to those obtained with the baseline prior (BLP). This outcome confirms that the HOD-informed priors effectively reduce unphysical parameter volume, mitigating projection effects and leading to a marginalized posterior that better recovers the underlying cosmological model.

\section{Data Availability}
The data used in this analysis are public along with Data Release 1 (details in \url{https://data.desi.lbl.gov/doc/releases/}). 
The data points corresponding to the figures from this paper is available at \url{https://doi.org/10.5281/zenodo.15178357}.

\section{Author Affiliations}
\label{sec:affiliations}
\input{affiliation}

\acknowledgments
AR acknowledges support from the Swiss National Science Foundation (SNF) ”Cosmology with 3D Maps of the Universe” research grant 200020 207379.
WP acknowledges the support of the Natural Sciences and Engineering Research Council of Canada (NSERC), [funding reference number RGPIN-2019-03908] and from the Canadian Space Agency.  
Research at Perimeter Institute is supported in part by the Government of Canada through the Department of Innovation, Science and Economic Development Canada and by the Province of Ontario through the Ministry of Colleges and Universities.
This research was enabled in part by support provided by Compute Ontario (computeontario.ca) and the Digital Research Alliance of Canada (alliancecan.ca).

This material is based upon work supported by the U.S. Department of Energy (DOE), Office of Science, Office of High-Energy Physics, under Contract No. DE–AC02–05CH11231, and by the National Energy Research Scientific Computing Center, a DOE Office of Science User Facility under the same contract. Additional support for DESI was provided by the U.S. National Science Foundation (NSF), Division of Astronomical Sciences under Contract No. AST-0950945 to the NSF’s National Optical-Infrared Astronomy Research Laboratory; the Science and Technology Facilities Council of the United Kingdom; the Gordon and Betty Moore Foundation; the Heising-Simons Foundation; the French Alternative Energies and Atomic Energy Commission (CEA); the National Council of Humanities, Science and Technology of Mexico (CONAHCYT); the Ministry of Science, Innovation and Universities of Spain (MICIU/AEI/10.13039/501100011033), and by the DESI Member Institutions: \url{https://www.desi.lbl.gov/collaborating-institutions}. Any opinions, findings, and conclusions or recommendations expressed in this material are those of the author(s) and do not necessarily reflect the views of the U. S. National Science Foundation, the U. S. Department of Energy, or any of the listed funding agencies.

The authors are honored to be permitted to conduct scientific research on I'oligam Du'ag (Kitt Peak), a mountain with particular significance to the Tohono O’odham Nation.

The authors acknowledge the use of the NASA astrophysics data system \url{https://ui.adsabs.harvard.edu} and the arXiv open-access repository \url{https://arxiv.org}. The software was hosted on the GitHub platform \url{https://github.com}. The manuscript was typeset using the overleaf cloud-based LaTeX editor \url{https://www.overleaf.com}.



\bibliographystyle{JHEP}
\bibliography{references_fixed,placeholder}

\end{document}

%% file: author.tex
\author[1,2]{H.~Zhang\orcidlink{0000-0001-6847-5254},}
\author[1,2,3]{M.~Bonici,}
\author[4]{A.~Rocher\orcidlink{0000-0003-4349-6424},}
\author[1,2,3]{W.~J.~Percival\orcidlink{0000-0002-0644-5727},}
\author[5]{A.~de~Mattia\orcidlink{0000-0003-0920-2947},}
\author[6]{J.~Aguilar,}
\author[7]{S.~Ahlen\orcidlink{0000-0001-6098-7247},}
\author[8]{O.~Alves,}
\author[9,10]{A.~Aviles\orcidlink{0000-0001-5998-3986},}
\author[6,11]{A.~Baleato Lizancos\orcidlink{0000-0002-0232-6480},}
\author[12,13]{D.~Bianchi\orcidlink{0000-0001-9712-0006},}
\author[14]{D.~Brooks,}
\author[6]{A.~Cuceu\orcidlink{0000-0002-2169-0595},}
\author[15]{A.~de la Macorra\orcidlink{0000-0002-1769-1640},}
\author[14]{P.~Doel,}
\author[6,11]{S.~Ferraro\orcidlink{0000-0003-4992-7854},}
\author[16]{N.~Findlay\orcidlink{0009-0007-0716-3477},}
\author[17]{A.~Font-Ribera\orcidlink{0000-0002-3033-7312},}
\author[4]{D.~Forero-Sánchez\orcidlink{0000-0001-5957-332X},}
\author[18,19]{J.~E.~Forero-Romero\orcidlink{0000-0002-2890-3725},}
\author[6]{S.~Gontcho A Gontcho\orcidlink{0000-0003-3142-233X},}
\author[20]{G.~Gutierrez,}
\author[21]{C.~Hahn\orcidlink{0000-0003-1197-0902},}
\author[22]{C.~Howlett\orcidlink{0000-0002-1081-9410},}
\author[23]{M.~Ishak\orcidlink{0000-0002-6024-466X},}
\author[11]{M.~Karamanis\orcidlink{0000-0001-9489-4612},}
\author[24]{R.~Kehoe,}
\author[25]{D.~Kirkby\orcidlink{0000-0002-8828-5463},}
\author[6]{A.~Kremin\orcidlink{0000-0001-6356-7424},}
\author[14]{O.~Lahav,}
\author[22]{Y.~Lai,}
\author[6]{M.~Landriau\orcidlink{0000-0003-1838-8528},}
\author[26]{L.~Le~Guillou\orcidlink{0000-0001-7178-8868},}
\author[6]{M.~E.~Levi\orcidlink{0000-0003-1887-1018},}
\author[17,27]{M.~Manera\orcidlink{0000-0003-4962-8934},}
\author[11]{M.~Maus\orcidlink{0000-0002-9020-911X},}
\author[28]{A.~Meisner\orcidlink{0000-0002-1125-7384},}
\author[17,29]{R.~Miquel,}
\author[1,2]{J.~Morawetz,}
\author[30]{J.~Moustakas\orcidlink{0000-0002-2733-4559},}
\author[16]{S.~Nadathur\orcidlink{0000-0001-9070-3102},}
\author[31]{J.~ A.~Newman\orcidlink{0000-0001-8684-2222},}
\author[9,32]{G.~Niz\orcidlink{0000-0002-1544-8946},}
\author[10,15]{H.~E.~Noriega\orcidlink{0000-0002-3397-3998},}
\author[5,6]{N.~Palanque-Delabrouille\orcidlink{0000-0003-3188-784X},}
\author[5]{M.~Pinon\orcidlink{0009-0009-3228-7126},}
\author[33]{F.~Prada\orcidlink{0000-0001-7145-8674},}
\author[34]{I.~P\'erez-R\`afols\orcidlink{0000-0001-6979-0125},}
\author[35]{G.~Rossi,}
\author[36,37]{S.~Saito\orcidlink{0000-0002-6186-5476},}
\author[38,39,40]{L.~Samushia\orcidlink{0000-0002-1609-5687},}
\author[41]{E.~Sanchez\orcidlink{0000-0002-9646-8198},}
\author[6]{D.~Schlegel,}
\author[8,42]{M.~Schubnell,}
\author[43]{H.~Seo\orcidlink{0000-0002-6588-3508},}
\author[28]{D.~Sprayberry,}
\author[8]{G.~Tarl\'{e}\orcidlink{0000-0003-1704-0781},}
\author[28]{B.~A.~Weaver,}
\author[16,44]{R.~Zhao\orcidlink{0000-0002-7284-7265},}
\author[6]{R.~Zhou\orcidlink{0000-0001-5381-4372},}

%% file: Tables/configs.tex


\begin{table}
\centering
\renewcommand{\arraystretch}{1.1}
\begin{tabular}{|l|cccccc|}
\hhline{|=======|}
\textbf{Quantity} & \texttt{BGS} & \texttt{LRG1} & \texttt{LRG2} & \texttt{LRG3} & \texttt{ELG2} & \texttt{QSO} \\
\hline
$z_{\rm eff}$ & 0.295 & 0.510 & 0.706 & 0.919 & 1.317 & 1.491 \\
$\sigma^2_n$ $[\hinvMpccubed]$ & 5723 & 5082 & 5229 & 9574 & 10692 & 47377 \\
$f_{\rm sat}$ & 0.15 & 0.15 & 0.15 & 0.15 & 0.10 & 0.03 \\
$\sigma_v$ $[\hinvmpc]$ & 5.06 & 6.20 & 6.20 & 6.20 & 3.11 & 5.68 \\
\hline
\end{tabular}
\caption{Relevant quantities used for basis conversion for each DESI tracer, listing the effective redshift ($z_{\rm eff}$), Poisson shot noise amplitude ($\sigma^2_n$ in $\hinvMpccubed$), satellite fraction ($f_{\rm sat}$), and characteristic velocity ($\sigma_v$ in $\hinvmpc$).}
\label{tab:config}
\end{table}

%% file: Tables/hoddist.tex
\begin{table}
\centering
\resizebox{\columnwidth}{!}{%
    \small 
\renewcommand{\arraystretch}{1.1}
    \begin{tabular}{|l|c|ccc|c|c|}
    \hhline{|=======|}
    \textbf{Parameters} & \texttt{BGS} & \texttt{LRG1} & \texttt{LRG2} & \texttt{LRG3} & \texttt{ELG2} & \texttt{QSO} \\
    \hline
    $\log_{10} M_{\mathrm{cut}}$ & \truncnorm{12}{14}{13}{1}&\multicolumn{3}{|c|}{\truncnorm{12}{13.8}{13}{1}}&\truncnorm{11.2}{13}{11.7}{0.5}&\truncnorm{11.2}{14}{12.7}{1}\\
    $\log_{10} M_1$ &  \truncnorm{12.5}{15.5}{14}{1} &\multicolumn{3}{|c|}{\truncnorm{12.5}{15.5}{14}{1}}&\truncnorm{12.5}{22}{18}{2.5}&\truncnorm{12}{16}{15}{1}\\ 
    $\sigma$ & \truncnorm{0}{3}{0.5}{0.5} &\multicolumn{3}{|c|}{\truncnorm{0}{3}{0.5}{0.5}}&\truncnorm{0.1}{3}{1}{1}&\truncnorm{0}{3}{0.5}{0.5}\\ 
    $\alpha$ & \truncnorm{0}{2}{1}{0.5} &\multicolumn{3}{|c|}{\truncnorm{0}{2}{1}{0.5}}&\truncnorm{0}{2}{1}{0.5}&\truncnorm{0.3}{2}{1}{0.5}\\ 
    $\kappa$ & \truncnorm{0}{3}{0.5}{0.5} &\multicolumn{3}{|c|}{\truncnorm{0}{3}{0.5}{0.5}}&\truncnorm{0}{3}{0.5}{1}&\truncnorm{0.3}{3}{0.5}{0.5}\\ 
    $\alpha_{\rm{cen}}$ & \truncnorm{0}{2}{1}{1} &\multicolumn{3}{|c|}{\truncnorm{0}{1}{0.4}{0.4}}&\truncnorm{0}{2}{0.5}{1}&\truncnorm{0}{2}{1.5}{1}\\
    $\alpha_{\rm{sat}}$ & \truncnorm{0}{2}{1}{1} &\multicolumn{3}{|c|}{\truncnorm{0}{2}{0.8}{0.4}}&\truncnorm{0}{2}{1.5}{1}&\truncnorm{0}{2}{0.2}{1}\\
    \hline
    $p_{\rm max}$ & --- &\multicolumn{3}{|c|}{---}&\truncnorm{0.02}{1}{0.1}{0.4}&---\\
    $\gamma$ & --- &\multicolumn{3}{|c|}{---}&\truncnorm{1}{15}{9}{5}&---\\
    $\log_{10} M_{\rm 1EE}$ & --- &\multicolumn{3}{|c|}{---}&\truncnorm{12.5}{18}{15}{2.5}&---\\
    \hline
    $z_{\rm sim}$ &0.3&0.5&0.725&0.95&1.325&1.475\\
    \hline
    \end{tabular}
}
\caption{Distributions of the HOD parameters used to calibrate the HOD-informed priors for each DESI tracer. Each cell indicates a truncated normal distribution, where the parameter is truncated to the interval shown in square brackets and has a mean and width specified in parentheses. The bottom row $z_\mathrm{sim}$ lists the simulation redshift used for each tracer.}
\label{tab:hoddist}
\end{table}

%% file: Tables/likelihoods.tex
\begin{table}
\centering
\resizebox{\columnwidth}{!}{%
    \small 
\renewcommand{\arraystretch}{1.1}
    \begin{tabular}{|l|ll|}
    \hhline{|===|}
    \textbf{Name} & \textbf{Description} & \textbf{Ref}\\
    \hline
    DESI-FS    & DESI DR1 full-shape likelihood with HOD-informed prior&\cite{DESICollaboration2024DESIQuasarsb,DESICollaboration2024DESIMeasurements}\\
    DESI-BAO   & DESI DR1 BAO reconstructed likelihood& \cite{DESICollaboration2024DESIQuasars,Adame2025DESIOscillations,Adame2025DESIForest}\\
    DESI       & Combined DESI DR1 FS$+$BAO likelihood with HOD-informed prior& this work\\
    \hline
    CMB-nl     & \textit{Planck} PR3 (\texttt{plik}$+$\texttt{simall}$+$\texttt{Commander}) TT, TE, EE, and lowE spectra&\cite{Collaboration2020PlanckParameters}\\
    CMB        & CMB-nl combined with \textit{Planck} PR4 (\texttt{NPIPE})$+$ACT lensing spectra&\cite{Carron2022CMBMaps,Madhavacheril2024TheParameters,Qu2024TheGrowth,MacCrann2024TheAnalysis}\\
    CMB-PR4    & \textit{Planck} PR4 (\texttt{HiLLiPoP}$+$\texttt{LoLLiPoP}$+$\texttt{Commander}) with \textit{Planck}$+$ACT lensing&\cite{Tristram2021PlanckRatio,Tristram2024CosmologicalPR4}\\
    \hline
    PantheonPlus &  SN Ia likelihood from PantheonPlus compilation&\cite{Brout2022TheConstraints,Scolnic2022TheRelease}\\
    Union3     & SN Ia likelihood from Union 3 compilation&\cite{Rubin2023UnionFramework}\\
    DES-SN5YR  & SN Ia likelihood of Year 5 supernova analysis from DES&\cite{Collaboration2024TheSet}\\
    \hline
    BBN        & Prior on $\Omega_\mathrm{b}h^2$ from Big Bang Nucleosynthesis, $\Omega_\mathrm{b}h^2\sim\mathcal{N}(0.02218,0.00055^2)$&\cite{Schoneberg2024TheUpdate}\\
    $n_{\mathrm{s10}}$ & Prior on $n_{\mathrm{s}}$ with width 10 times wider than \textit{Planck}, $n_{\mathrm{s}}\sim\mathcal{N}(0.9649,0.042^2)$&\cite{Collaboration2020PlanckParameters}\\
    \hline
    \end{tabular}
}
\caption{Summary of likelihoods used in this analysis. The first column lists the shorthand notation for each likelihood, followed by a brief description and relevant references.}
\label{tab:likelihoods}
\end{table}

%% file: Tables/results.tex
\begin{table}
\centering
\resizebox{\columnwidth}{!}{%
    \small 
\setcellgapes{3pt}\makegapedcells  
\renewcommand{\arraystretch}{2.1}
    \begin{tabular}{|l|cccccc|}
    \hhline{|=======|}
    \bf{Model/Dataset} & $\Omega_{\mathrm{m}}$ & $\sigma_8$ & $S_8$ & \makecell[c]{$H_0$\\[0.1cm] [${\rm km\,s^{-1} Mpc^{-1}}$]} & $w_0$& $w_a$\\[0.1cm]
    \hline
    {\bf Flat} $\boldsymbol{\Lambda}${\bf CDM} &&&&&&\\
    \makecell[l]{DESI-FS\\+BBN+$n_{\mathrm{s10}}$} & $0.2874^{+0.0094}_{-0.010}$ & $0.832^{+0.024}_{-0.028}$ & $0.814^{+0.027}_{-0.031}$ & $69.9\pm 1.0$ & --- & --- \\
    \makecell[l]{DESI\\+BBN+$n_{\mathrm{s10}}$} & $0.2994\pm 0.0090$ & $0.836^{+0.024}_{-0.027}$ & $0.835^{+0.027}_{-0.030}$ & $68.40\pm 0.73$ & --- & --- \\
    \hdashline
    DESI+CMB-nl & $0.3065\pm 0.0052$ & $0.8098\pm 0.0070$ & $0.818\pm 0.011$ & $68.00\pm 0.39$ & --- & --- \\
    DESI+CMB & $0.3072\pm 0.0049$ & $0.8123\pm 0.0052$ & $0.8220\pm 0.0089$ & $67.95\pm 0.37$ & --- & --- \\
    DESI+CMB-PR4 & $0.3062\pm 0.0047$ & $0.8119\pm 0.0045$ & $0.8201\pm 0.0086$ & $67.92\pm 0.36$ & --- & --- \\
    {\bf Flat} $\boldsymbol{w_0 w_a}${\bf CDM} &&&&&&\\
    \hline
    \makecell[l]{DESI-FS\\+BBN+$n_{\mathrm{s10}}$} & $0.301\pm 0.026$ & $0.799^{+0.031}_{-0.039}$ & $0.799\pm 0.038$ & $68.9^{+2.7}_{-3.5}$ & $-0.74^{+0.30}_{-0.25}$ & $-1.11\pm 0.88$ \\
    \makecell[l]{DESI\\+BBN+$n_{\mathrm{s10}}$} & $0.324\pm 0.025$ & $0.804^{+0.030}_{-0.035}$ & $0.834\pm 0.034$ & $66.6^{+2.2}_{-2.8}$ & $-0.70\pm 0.25$ & $-1.10\pm 0.79$ \\
    \hdashline
    DESI+CMB-nl & $0.331\pm 0.021$ & $0.804^{+0.017}_{-0.019}$ & $0.844\pm 0.015$ & $65.8^{+1.9}_{-2.2}$ & $-0.62\pm 0.21$ & $-1.28^{+0.63}_{-0.57}$ \\
    DESI+CMB & $0.330\pm 0.021$ & $0.803^{+0.016}_{-0.019}$ & $0.841\pm 0.012$ & $65.9^{+1.8}_{-2.2}$ & $-0.63\pm 0.21$ & $-1.24^{+0.61}_{-0.55}$ \\
    DESI+CMB-PR4 & $0.326\pm 0.021$ & $0.804^{+0.016}_{-0.019}$ & $0.837\pm 0.013$ & $66.2^{+1.8}_{-2.3}$ & $-0.68\pm 0.21$ & $-1.08\pm 0.58$ \\
    \hdashline
    \makecell[l]{DESI+CMB\\+PantheonPlus} & $0.3110\pm 0.0064$ & $0.8172\pm 0.0084$ & $0.8319\pm 0.0090$ & $67.79\pm 0.65$ & $-0.844\pm 0.058$ & $-0.66^{+0.25}_{-0.22}$ \\
    \makecell[l]{DESI+CMB\\+Union3} & $0.3245\pm 0.0091$ & $0.8067\pm 0.0096$ & $0.8388\pm 0.0095$ & $66.41\pm 0.89$ & $-0.685\pm 0.092$ & $-1.11^{+0.34}_{-0.30}$ \\
    \makecell[l]{DESI+CMB\\+DES-SN5YR} & $0.3185\pm 0.0062$ & $0.8115\pm 0.0081$ & $0.8361\pm 0.0090$ & $67.02\pm 0.60$ & $-0.752\pm 0.062$ & $-0.93^{+0.27}_{-0.24}$ \\
    \hline
    \end{tabular}
}
\caption{68\% credible intervals for the cosmological parameters $\Omega_m$, $\sigma_8$, $S_8$, $H_0$, $w_0$, and $w_a$ under various data and model combinations. The top section lists results for the flat $\Lambda$CDM model, while the bottom section shows the flat $w_0w_a$CDM extension. Each row corresponds to a different combination of datasets, as indicated in the first column.
}

\label{tab:results}
\end{table}

%% file: affiliation.tex
\noindent \hangindent=.5cm $^{1}${Department of Physics and Astronomy, University of Waterloo, 200 University Ave W, Waterloo, ON N2L 3G1, Canada}

\noindent \hangindent=.5cm $^{2}${Waterloo Centre for Astrophysics, University of Waterloo, 200 University Ave W, Waterloo, ON N2L 3G1, Canada}

\noindent \hangindent=.5cm $^{3}${Perimeter Institute for Theoretical Physics, 31 Caroline St. North, Waterloo, ON N2L 2Y5, Canada}

\noindent \hangindent=.5cm $^{4}${Institute of Physics, Laboratory of Astrophysics, \'{E}cole Polytechnique F\'{e}d\'{e}rale de Lausanne (EPFL), Observatoire de Sauverny, Chemin Pegasi 51, CH-1290 Versoix, Switzerland}

\noindent \hangindent=.5cm $^{5}${IRFU, CEA, Universit\'{e} Paris-Saclay, F-91191 Gif-sur-Yvette, France}

\noindent \hangindent=.5cm $^{6}${Lawrence Berkeley National Laboratory, 1 Cyclotron Road, Berkeley, CA 94720, USA}

\noindent \hangindent=.5cm $^{7}${Physics Dept., Boston University, 590 Commonwealth Avenue, Boston, MA 02215, USA}

\noindent \hangindent=.5cm $^{8}${University of Michigan, 500 S. State Street, Ann Arbor, MI 48109, USA}

\noindent \hangindent=.5cm $^{9}${Instituto Avanzado de Cosmolog\'{\i}a A.~C., San Marcos 11 - Atenas 202. Magdalena Contreras. Ciudad de M\'{e}xico C.~P.~10720, M\'{e}xico}

\noindent \hangindent=.5cm $^{10}${Instituto de Ciencias F\'{\i}sicas, Universidad Nacional Aut\'onoma de M\'exico, Av. Universidad s/n, Cuernavaca, Morelos, C.~P.~62210, M\'exico}

\noindent \hangindent=.5cm $^{11}${University of California, Berkeley, 110 Sproul Hall \#5800 Berkeley, CA 94720, USA}

\noindent \hangindent=.5cm $^{12}${Dipartimento di Fisica ``Aldo Pontremoli'', Universit\`a degli Studi di Milano, Via Celoria 16, I-20133 Milano, Italy}

\noindent \hangindent=.5cm $^{13}${INAF-Osservatorio Astronomico di Brera, Via Brera 28, 20122 Milano, Italy}

\noindent \hangindent=.5cm $^{14}${Department of Physics \& Astronomy, University College London, Gower Street, London, WC1E 6BT, UK}

\noindent \hangindent=.5cm $^{15}${Instituto de F\'{\i}sica, Universidad Nacional Aut\'{o}noma de M\'{e}xico,  Circuito de la Investigaci\'{o}n Cient\'{\i}fica, Ciudad Universitaria, Cd. de M\'{e}xico  C.~P.~04510,  M\'{e}xico}

\noindent \hangindent=.5cm $^{16}${Institute of Cosmology and Gravitation, University of Portsmouth, Dennis Sciama Building, Portsmouth, PO1 3FX, UK}

\noindent \hangindent=.5cm $^{17}${Institut de F\'{i}sica d’Altes Energies (IFAE), The Barcelona Institute of Science and Technology, Edifici Cn, Campus UAB, 08193, Bellaterra (Barcelona), Spain}

\noindent \hangindent=.5cm $^{18}${Departamento de F\'isica, Universidad de los Andes, Cra. 1 No. 18A-10, Edificio Ip, CP 111711, Bogot\'a, Colombia}

\noindent \hangindent=.5cm $^{19}${Observatorio Astron\'omico, Universidad de los Andes, Cra. 1 No. 18A-10, Edificio H, CP 111711 Bogot\'a, Colombia}

\noindent \hangindent=.5cm $^{20}${Fermi National Accelerator Laboratory, PO Box 500, Batavia, IL 60510, USA}

\noindent \hangindent=.5cm $^{21}${Steward Observatory, University of Arizona, 933 N. Cherry Avenue, Tucson, AZ 85721, USA}

\noindent \hangindent=.5cm $^{22}${School of Mathematics and Physics, University of Queensland, Brisbane, QLD 4072, Australia}

\noindent \hangindent=.5cm $^{23}${Department of Physics, The University of Texas at Dallas, 800 W. Campbell Rd., Richardson, TX 75080, USA}

\noindent \hangindent=.5cm $^{24}${Department of Physics, Southern Methodist University, 3215 Daniel Avenue, Dallas, TX 75275, USA}

\noindent \hangindent=.5cm $^{25}${Department of Physics and Astronomy, University of California, Irvine, 92697, USA}

\noindent \hangindent=.5cm $^{26}${Sorbonne Universit\'{e}, CNRS/IN2P3, Laboratoire de Physique Nucl\'{e}aire et de Hautes Energies (LPNHE), FR-75005 Paris, France}

\noindent \hangindent=.5cm $^{27}${Departament de F\'{i}sica, Serra H\'{u}nter, Universitat Aut\`{o}noma de Barcelona, 08193 Bellaterra (Barcelona), Spain}

\noindent \hangindent=.5cm $^{28}${NSF NOIRLab, 950 N. Cherry Ave., Tucson, AZ 85719, USA}

\noindent \hangindent=.5cm $^{29}${Instituci\'{o} Catalana de Recerca i Estudis Avan\c{c}ats, Passeig de Llu\'{\i}s Companys, 23, 08010 Barcelona, Spain}

\noindent \hangindent=.5cm $^{30}${Department of Physics and Astronomy, Siena College, 515 Loudon Road, Loudonville, NY 12211, USA}

\noindent \hangindent=.5cm $^{31}${Department of Physics \& Astronomy and Pittsburgh Particle Physics, Astrophysics, and Cosmology Center (PITT PACC), University of Pittsburgh, 3941 O'Hara Street, Pittsburgh, PA 15260, USA}

\noindent \hangindent=.5cm $^{32}${Departamento de F\'{\i}sica, DCI-Campus Le\'{o}n, Universidad de Guanajuato, Loma del Bosque 103, Le\'{o}n, Guanajuato C.~P.~37150, M\'{e}xico}

\noindent \hangindent=.5cm $^{33}${Instituto de Astrof\'{i}sica de Andaluc\'{i}a (CSIC), Glorieta de la Astronom\'{i}a, s/n, E-18008 Granada, Spain}

\noindent \hangindent=.5cm $^{34}${Departament de F\'isica, EEBE, Universitat Polit\`ecnica de Catalunya, c/Eduard Maristany 10, 08930 Barcelona, Spain}

\noindent \hangindent=.5cm $^{35}${Department of Physics and Astronomy, Sejong University, 209 Neungdong-ro, Gwangjin-gu, Seoul 05006, Republic of Korea}

\noindent \hangindent=.5cm $^{36}${Institute for Multi-messenger Astrophysics and Cosmology, Department of Physics, Missouri University of Science and Technology, 1315 N Pine St, Rolla, MO 65409 U.S.A.}

\noindent \hangindent=.5cm $^{37}${Kavli Institute for the Physics \& Mathematics of the Universe, University of Tokyo, Kashiwa, Chiba 227-8583, Japan}

\noindent \hangindent=.5cm $^{38}${Abastumani Astrophysical Observatory, Tbilisi, GE-0179, Georgia}

\noindent \hangindent=.5cm $^{39}${Department of Physics, Kansas State University, 116 Cardwell Hall, Manhattan, KS 66506, USA}

\noindent \hangindent=.5cm $^{40}${Faculty of Natural Sciences and Medicine, Ilia State University, 0194 Tbilisi, Georgia}

\noindent \hangindent=.5cm $^{41}${CIEMAT, Avenida Complutense 40, E-28040 Madrid, Spain}

\noindent \hangindent=.5cm $^{42}${Department of Physics, University of Michigan, 450 Church Street, Ann Arbor, MI 48109, USA}

\noindent \hangindent=.5cm $^{43}${Department of Physics \& Astronomy, Ohio University, 139 University Terrace, Athens, OH 45701, USA}

\noindent \hangindent=.5cm $^{44}${National Astronomical Observatories, Chinese Academy of Sciences, A20 Datun Road, Chaoyang District, Beijing, 100101, P.~R.~China}